\documentclass[showpacs,amsmath,amssymb,aps,twocolumn,superscriptaddress,prx]{revtex4-2} 
\usepackage{graphicx}
\usepackage{amsthm,amssymb,amsmath,braket,mathdots}
\usepackage{bm}
\usepackage[pagebackref=false,pdfnewwindow=true]{hyperref} 
\usepackage{epstopdf,psfrag}
\usepackage{relsize,amsbsy}
\usepackage[export]{adjustbox}
\usepackage{makecell}
\usepackage{graphicx,xcolor,tikz}
\usepackage{float}
\usepackage{mathtools}
\usepackage[centerlast]{caption}
\usepackage{hyperref}

\newcommand{\be}{\begin{equation}}
\newcommand{\ee}{\end{equation}}

\newcommand{\bit}{\begin{enumerate}}
	\newcommand{\eit}{\end{enumerate}}

\definecolor{bananayellow}{rgb}{1.0, 0.88, 0.21}
\definecolor{straw}{rgb}{0.32, 0.28, 0.1}

\begin{document}
	\title{Prethermalization in aperiodically kicked many-body dynamics
 }
		\author{Jin Yan}
\affiliation{\small Max-Planck-Institut f{\"u}r Physik komplexer Systeme, N{\"o}thnitzer Stra{\ss}e 38, 01187 Dresden, Germany}

	\author{Roderich Moessner}
		\affiliation{\small Max-Planck-Institut f{\"u}r Physik komplexer Systeme, N{\"o}thnitzer Stra{\ss}e 38, 01187 Dresden, Germany}

		\author{Hongzheng Zhao}
\affiliation{\small Max-Planck-Institut f{\"u}r Physik komplexer Systeme, N{\"o}thnitzer Stra{\ss}e 38, 01187 Dresden, Germany}

\begin{abstract}
Driven many-body systems typically experience heating  due to the lack of energy conservation. Heating  may be  suppressed for time-periodic drives, but little is known for less regular drive protocols.
In this work, we investigate the heating dynamics in aperiodically kicked systems, specifically those driven by quasi-periodic Thue-Morse or a family of {\it random} sequences with $n$-multipolar temporal correlations. We demonstrate that multiple heating channels can be eliminated {\it even away from the high-frequency regime}. The number of eliminated channels increases with multipolar order $n$. We illustrate this in a classical kicked rotor chain where we find  a long-lived prethermal regime. When the static Hamiltonian only involves the kinetic energy, the prethermal lifetime $t^*$ can strongly depend on the temporal correlations of the drive, with a  power-law dependence on the {\it kick  strength} $t^*\sim K^{-2n}$, for which we can account  using a simple linearization argument.
\end{abstract}
	\maketitle 
 
	\let\oldaddcontentsline\addcontentsline 
	\renewcommand{\addcontentsline}[3]{} 

 
\textit{Introduction.---} Time-dependent many-body systems have attracted sustained interest due to their ubiquity in nature and the potential to realize novel non-equilibrium phases of matter. One typical example is the discrete time crystal which spontaneously breaks discrete time translation symmetry (TTS)~\cite{khemani2019brief,zaletel2023colloquium}. However, due to the absence of energy conservation, closed driven systems tend to heat up and lose any {non-trivial correlations}~\cite{d2014long,lazarides2014equilibrium}. Therefore, understanding and controlling the onset of heating in time-dependent systems is key to their stabilization, and to the realization of exotic non-equilibrium phenomena.

While heating generally occurs in time-dependent many-body systems, it can be parametrically suppressed, e.g., by using high-frequency drives ~\cite{bukov2015universal,abanin2015exponentially,kuwahara2016floquet,weidinger2017floquet,sieberer2019digital,rubio2020floquet,hodson2021energy,mcroberts2022prethermalization,beatrez2023critical,ho2023quantum} or by using weak drive amplitudes~\cite{vajna2018replica,fleckenstein2021thermalization,mori2022heating} in periodically driven (Floquet) systems.  Examples include spin systems with a bounded local energy scale, where an exponentially long-lived prethermal regime appears before heating takes over~\cite{mori2018floquet,haldar2018onset,vajna2018replica,howell2019asymptotic,pizzi2021classical,ikeda2021fermi,ye2021floquet,jin2023fractionalized}. {A similar prethermal phenomenon can also manifest in
kicked systems which have been extensively studied in the context of digital quantum simulation~\cite{sieberer2019digital,lysne2020small,joshi2022probing} and the fundamental discussion of chaos~\cite{chirikov1971research,haake1987classical,kaneko1989diffusion, konishi1990diffusion, falcioni1991ergodic, mulansky2011strong}. A paradigmatic example is the interacting kicked rotor, where heating takes the form of Arnold diffusion ~\cite{chirikov1997arnold,richter2014visualization,citro2015dynamical,rozenbaum2017dynamical,lellouch2020dynamics,kundu2021dynamics,haldar2021prethermalization,russomanno2021chaos,martinez2022low,cao2022interaction}:
before their eventual diffusive dynamics with unbounded energy growth~\cite{kaneko1989diffusion}, heating only occurs with a probability exponentially small in the kick strength in the prethermal regime~\cite{rajak2019characterizations}.}

It is natural to ask: can heating be efficiently suppressed in many-body systems without TTS, e.g., when drives are quasi-periodic, or even random? This is a notoriously difficult question as breaking TTS generally opens up
further deleterious heating channels that can destabilize systems rapidly~\cite{wen2021periodically,timms2021quantized,pilatowskycameo2023complete}. {For certain piecewise constant and continuous quasi-periodic drives, this is known to be possible in the high-frequency regime~\cite{verdeny2016quasi,dumitrescu2018logarithmically,mukherjee2020restoring,lapierre2020fine,zhao2021random,cai20221,ying2022floquet,he2022quasi,long2022many,martin2022effect,zhao2022temporal,tiwari2023dynamical}. }Rigorous bounds on heating rates can also be established by generalizing the Floquet theory~\cite{else2020long,mori2021rigorous}. However, this becomes obscure for 
kicked systems as the high-frequency limit of {\it kicks} in principle allows a {\it divergent} rate of energy input into the system. 
Aperiodically kicked systems have been most limited to few-body settings~\cite{casati1989anderson,lemarie2009universality,goldfriend2020quasi,santhanam2022quantum,vuatelet2023dynamical} and it remains an outstanding challenge to control heating in the thermodynamic limit. 

Here, we give an affirmative answer by investigating many-body systems kicked by a family of structured binary random protocols known as random multipolar drives (RMD)~\cite{zhao2021random}. These drives exhibit a multipolar correlation indexed by a non-negative integer $n$: for $n=0$, the drive is purely random and generated from the binary options $\{s_0^+,s_0^-\}=\{+,-\}$; for $n=1$, it consists of a random sequence of two elementary {\it dipolar blocks}, $\{s_1^+,s_1^-\}=\{(-, +),(+, -)\}$; and the $n$th order multipolar blocks are recursively generated by concatenating two different $(n-1)$th order blocks, $\{s_{n}^{+},s_n^-\}=\{(s_{n-1}^{-},s_{n-1}^{+}),(s_{n-1}^{+},s_{n-1}^{-})\}$. In the $n\to\infty$ limit, $s_n^{\pm}$ produces the quasi-periodic Thue-Morse (TM) sequence~\cite{nandy2017aperiodically,mori2021rigorous}. RMD notably suppresses the low-frequency components in the driving spectrum and it suffices to reduce heating algebraically in the high-frequency regime~\cite{zhao2021random}.

In this work, instead of focusing on the high-frequency regime, we exploit the self-similarity inherent in the RMD sequence to demonstrate that heating can be parametrically controlled by the kick strength. Through a perturbative expansion, we derive an effective Hamiltonian that governs the initial time evolution. Remarkably, the self-similar multipolar structure leads to exact cancellations of numerous terms in the effective Hamiltonian, thereby eliminating the corresponding heating channels. This mechanism of heating suppression is independent of the specific model and is applicable to both quantum and classical many-body systems.

For numerical efficiency, 
we demonstrate this effect in a concrete model, namely a kicked chain of classical rotors. Starting from low-temperature initial states, the system exhibits a long-lived prethermal regime before heating up. The lifetime scaling depends on the microscopic details of the kicked system and if the static part only involves the kinetic energy,
the lifetime scales as a power law with a tunable exponent $2n$. This  we account for by analysing the linear stability of the system. In the quasi-periodic TM limit, we also show that the lifetime grows faster than
any power law but slower than exponentially. 

In the following, we first consider a general kicked system and demonstrate heating suppression in a perturbative expansion of the effective Hamiltonian in the small kick strength. We then present the results on the chain of rotors  before a concluding discussion.

\textit{Setting.---} 
Consider the time-dependent Hamiltonian $H(t) = H + V\Delta(t)$, where $H$ denotes the static part and $V$ defines the kick with $\Delta(t) = \sum_l K_l \delta(t - l \tau)$, and the kick strength $K_l$ with period $\tau$. We focus on the intermediate frequency regime, i.e. $\tau$ is not necessarily small.
Suppose the strength $K_l$ has binary choices $\pm K$ following an $n$-RMD sequence. For $n = 0$, there are two possible unitary time evolution operators:
\begin{equation}
\label{eq:elementary}
    U_0^{+} = e^{-i\tau H}e^{-iKV},\quad
    U_0^{-} = e^{-i\tau H}e^{iKV}.
\end{equation}

We can formally define the time-independent effective Hamiltonian $H_0^{\pm}$ through the relation $U_0^{\pm} = \exp(-i\tau H_0^{\pm})$. For weak kick strength $K$, we can perturbatively construct the effective Hamiltonian as $H^{\pm}_{0} = \sum_{m=0}^{\infty} K^m\Omega_{0,m}^{\pm}$, describing the dynamics at times $t = l\tau$. Although such an expansion may diverge for many-body systems, we expect that its truncation at low orders in $K$ will approximate the initial time evolution. The lowest order term is simply the static Hamiltonian, $\Omega_{0,0}^{\pm} = H$. 

Via the replica resummation of the Baker-Campbell-Hausdorff series, the leading correction can be expressed in a compact form
~\cite{vajna2018replica}
\begin{equation}
\begin{aligned}
\label{eq.expansion_n0}
&\Omega_{0,1}^+=-\Omega_{0,1}^-\coloneqq \Omega_{0,1}=\frac{-i {\mathrm{ad}_{H}} e^{-i \tau \mathrm{ad}_{H}}}{e^{-i \tau \mathrm{ad}_{H}}-1} V, 
\end{aligned}
\end{equation}
where $\operatorname{ad}_X(Y)=[X, Y]$ is the Lie derivative. It can also be expanded in a power series in $\tau$ as 
\begin{eqnarray}
\label{eq:Omega_expansion}
    \Omega_{0,1} = V\tau^{-1} -i [H,V]/2 + \mathcal{O}(\tau^1),
\end{eqnarray}
where higher order terms only contain nested commutators of the form
$
    [H,V]_s \coloneqq  [H,\dots,[H,V]\dots]
$
with a single kick $V$ but multiple ($s$) $H$ operators. Generally, $\Omega_{0,1}$ does not vanish, and initially, the time evolution is dominated by $H \pm K\Omega_{0,1}$. The term $\Omega_{0,1}$ occurs randomly with an amplitude linear in $K$, and one expects it to quickly destabilize the system and induce heating in a short time.

We now use the self-similar structure of RMD protocols to show that many terms of  order $\mathcal{O}(K)$ in the effective Hamiltonian can be eliminated. Furthermore, if the condition 
\begin{equation}
\label{eq:condition}
   [H,V]_s=0,  
\end{equation}
can be satisfied for $\forall s\ge n_c$ with some integer $n_c$, for $n-$RMD systems with any $n\geq n_c$, random perturbations start appearing at a higher order $\mathcal{O}(K^3)$ and hence heating can be significantly suppressed. 

To see this, we first observe that higher-order multipolar operators can be recursively obtained using the relation
\begin{eqnarray}
\label{eq:recursive_relation}
    U_{n}^{\pm} = U_{n-1}^{\mp}U_{n-1}^{\pm},
\end{eqnarray}
where $U_n^{\pm}$ generates the time evolution over the duration $2^n\tau$~\cite{zhao2021random}. For $n$-RMD systems, the time evolution is given by a random sequence of multipolar operators $U_n^{\pm}$. Similarly, the effective Hamiltonian $H_n^{\pm}$ is defined through $U_n^{\pm} = \exp(-i2^n\tau H_n^{\pm})$, governing the stroboscopic time evolution ($t = 2^n \tau l$ for integers $l$). The perturbative expansion is denoted as $H^{\pm}_{n} = \sum_{m=0}^{\infty} K^m\Omega_{n,m}^{\pm}$. Notably, the time evolution operators in Eq.~\ref{eq:elementary} possess the special property that $U_0^{+}$ can be mapped to $U_0^-$ by changing $K\to-K$. Thus, terms in the effective Hamiltonians coincide for even orders in $K$, while differing by a minus sign for odd orders, given by
\begin{equation}
\begin{aligned}
    \Omega_{n,m}^{+} = (-1)^m\Omega_{n,m}^- \coloneqq \Omega_{n,m}.
\end{aligned}
\label{eq.symmetry}
\end{equation}

Similar to the purely random drive ($n=0$), the initial stroboscopic time evolution is governed by $H \pm K\Omega_{n,1}$, and the system may still exhibit rapid heating. However, it is noteworthy that the self-similar construction in Eq.~\ref{eq:recursive_relation} and the symmetry property in Eq.~\ref{eq.symmetry} lead to an important observation: several terms in $\Omega_{n,1}$ actually vanish, resulting in the remarkable property
\begin{equation}
\begin{aligned}
    \Omega_{n,1} = \sum_{s=n}^{\infty}f_{n,s}\tau^s[H,\Omega_{0,1}]_s ,
\end{aligned}
    \label{eq.ansatz_main}
\end{equation}
where the summation starts from $s=n$, although obtaining the coefficient $f_{n,s}$ can be a cumbersome task.
Importantly, Eq.~\ref{eq.ansatz_main} suggests that, to the leading order of $\mathcal{O}(K)$, heating can only occur through heating channels in the form of $[H,V]_s$ with $s\geq n$, while all other heating channels are strictly forbidden. The derivation of this expression is presented in Sec.~\ref{sec:perturbative_expansion} of the Supplementary Materials (SM).
Now, if the condition given by Eq.~\ref{eq:condition} is satisfied, all terms in $\Omega_{n,1}$ vanish. Consequently, the stroboscopic time evolution of the system is effectively governed by the Hamiltonian $H^{\pm}_n = \bar{H}_n \pm \mathcal{O}(K^3)$, where the static part is denoted by $\bar{H}_n = H + K^2\Omega_{n,2}$.
Therefore, the RMD kicked systems first relax to a prethermal ensemble determined by $\bar{H}_n$ before notable heating is induced by random perturbations of order $\mathcal{O}(K^3)$.

Although we use the perturbative expansion for quantum systems, it is important to note that this mechanism of heating suppression equally applies to classical many-body systems. The Liouville equation, which describes the phase-space distribution of a classical system, exhibits a structural similarity to the Schrödinger equation in quantum systems. Consequently, the effective Hamiltonian for classical systems can be obtained by formally replacing the commutator $[\dots]/i$ in its quantum counterpart with the Poisson bracket $\{\dots \}$~\cite{mori2018floquet}. Due to the computational efficiency of numerical simulations for large classical systems, we proceed to demonstrate this heating suppression in a classical rotor system.

\textit{Many-body kicked rotors.---}
We consider many-body rotors with the static kinetic energy $(H=H_{\mathrm{kin}})$ and the kicked nearest-neighboring interactions $(V=V_{\mathrm{int}})$,  
\begin{eqnarray}
\label{eq:Hamiltonian_rotor}
        H_{\mathrm{kin}} = \frac{1}{2}\sum_j p_j^2,\ \  V_{\mathrm{int}} = \sum_j\cos \left(q_{j+1}-q_j\right),
\end{eqnarray} 
where $p_j$ and $q_j$ for $j = 1,...,N$ are the {conjugate} angular momenta and angles of $N$ rotors, respectively. Periodic boundary conditions are used ($q_1=q_{N+1}$). The interaction preserves the total angular momentum $\sum_{j=1}^N p_j$. 
 
\begin{figure}
\centering
\includegraphics[width=0.98\linewidth]{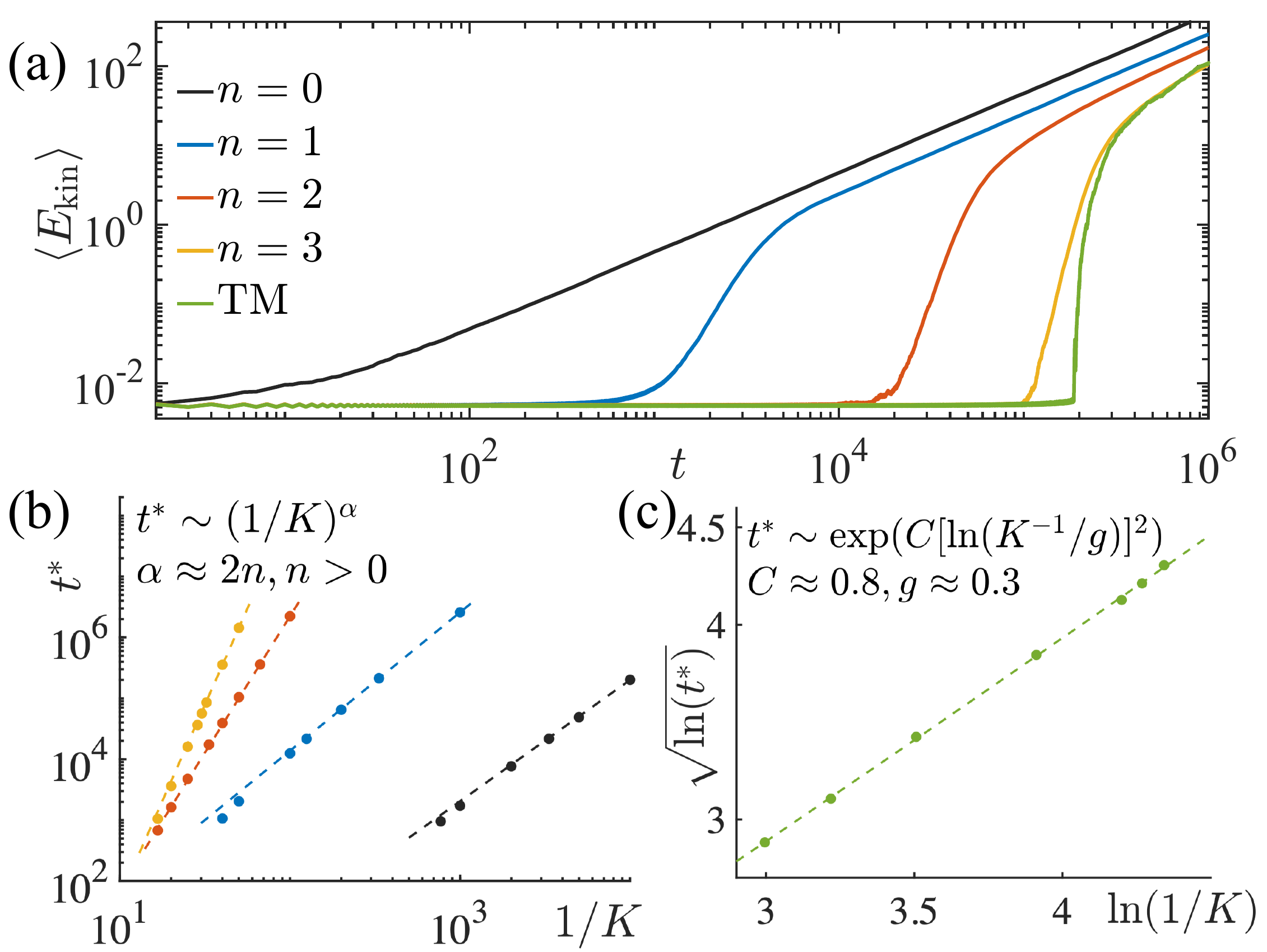}
\caption{(a) Time evolution of the averaged kinetic energy for $n-$RMD and Thue-Morse (TM) drive with $K=0.03$ in a log-log scale. (b) Dependence of the prethermal lifetime $t^*$ on $1/K$ in a log-log scale. Dashed lines ($K^{-2n}$ for $n>0$ and $K^{-2}$ for $n=0$) are plotted to guide the eyes.
(c) Prethermal lifetime $t^*$ scaling for TM drives.
}
\label{fig-B0}
\end{figure}
As we will demonstrate below, the condition in Eq.~\ref{eq:condition} can be approximately satisfied when the angular momentum distribution is narrow. In the prethermal regime, the width of the distribution is determined by the temperature $T$, which is controllably small and scales with the kick strength as $T\sim K^2$.

To begin, we derive the nested Poisson brackets $\{H,V\}_s$ for kicked rotors, which reduce to $\sum_j(p_j-p_{j+1})^s\sin(q_j-q_{j+1})$ for odd $s$ and $\sum_j(p_j-p_{j+1})^s\cos(q_j-q_{j+1})$ for even $s$. Therefore, for multipolar order $n\ge 1$, the expression in Eq.~\ref{eq.ansatz_main} implies that the dominant random perturbation $\Omega_{n,1}$ only contains terms proportional to $(p_j-p_{j+1})$ or its higher powers. These terms become negligible when the kinetic energy distribution is sufficiently narrow. The suppression becomes stronger with higher multipolar orders, and Eq.~\ref{eq:condition} can be more effectively satisfied for larger multipolar order $n$.

When we start from the initial condition $p_j=\tilde{p}$ for all $j$, Eq.~\ref{eq:condition} is fulfilled exactly, and the initial time evolution is governed by the static effective Hamiltonian $\bar{H}_n=H_{\mathrm{kin}}+K^2\Omega_{n,2}$. As this Hamiltonian is generally non-integrable, the angular momentum distribution spreads. In the prethermal regime, it approximately reaches the Gibbs distribution $\prod_{j=1}^N \exp \left[-{(p_j-\tilde{p})^2}/{2 T}\right]$~\cite{rajak2019characterizations}. The width of the distribution is determined by an effective prethermal temperature $T$. As shown in Sec.~\ref{sec.temperature}, this temperature can be controlled to be small for weak kicks ($T\sim K^2$).

\textit{Numerical simulation.---}
We now confirm the possibility of prethermalization through numerical simulations. The time evolution of RMD kicked rotors can be generated using a set of discretized classical equations of motion (EOM):
\begin{equation}
\begin{aligned}
p_j(t+1) &= p_j(t) \pm K \left[ \sin(q_{j+1}(t) - q_j(t)) \right. \\
& \quad \left. + \sin(q_{j-1}(t) - q_j(t)) \right], \\
q_j(t+1) &= q_j(t) + \tau p_j(t+1),\ \text{for}\ j = 1, 2, ..., N,
\end{aligned}
\label{eq-of-motion}
\end{equation}
where the $\pm$ sign follows the RMD sequence and $t$ labels the number of kicks. We choose $\tau=1$ for numerical simulations.

The spreading of the angular momentum distribution can be quantified by the kinetic energy density $E_{\text{kin}}(t) := \frac{1}{2N}\sum_{i=1}^N p_i^2(t)$, making it a suitable measure of heating and temperature increase. The initial conditions are chosen such that the angles $q_j$ uniformly distribute between $0$ and $2\pi$, and the angular momentum $p_j = 0.1$ is fixed for all rotors, satisfying the condition in Eq.~\ref{eq:condition}.
In Fig.~\ref{fig-B0}(a), we depict the time evolution of the averaged kinetic energy $\langle E_{\text{kin}} \rangle$, averaged over $350$ noise realizations with different initial states, for a fixed kicking strength $K=0.03$ and rotor number $N=500$~\footnote{$N=500$ is chosen to sufficiently mimic the heating behavior in thermodynamically large systems; further details can be found in Sec.~\ref{sec.finite_size}.}. For multipolar order $n\geq 1$, the averaged kinetic energy remains almost unchanged for a long timescale $t^*$. However, as the kinetic energy is unbounded, it eventually increases when heating takes over. We observe that for larger $n$, the timescale $t^*$ remarkably extends by several orders of magnitude, reaching its largest value in the quasi-periodic TM limit. In contrast, for the fully random drive ($n=0$), unbounded diffusion starts even at very early times, and no prethermal regime can be established.

We quantify the prethermal lifetime $t^*$ and its dependence on the kicking strength $K$. To extract $t^*$, one can fit the averaged kinetic energy up to time $t_f$ with a power law $t^b$ and monitor the power $b$ for different $t_f$~\cite{rajak2019characterizations}. During the prethermal regime, the power $b$ remains close to zero, and $t^*$ is determined when $b$ first reaches a threshold. In our numerical simulations, we choose $b = 0.05$, but our findings are independent of the specific threshold value as long as it is small.

Fig.~\ref{fig-B0}(b) illustrates the dependence of $t^*$ on the kick strength for different multipolar orders. Using a log-log scale, a linear fit suggests that the prethermal lifetime follows an algebraic dependence on the kick strength, $t^*\sim (1/K)^{\alpha}$. The scaling exponent ${\alpha}$ can be determined through numerical fitting. For $n=0$, the exponent is close to 2. It remains approximately the same for $n=1$, although the prefactor differs by three orders of magnitude, indicating significant suppression of heating due to the dipolar structure. Interestingly, for higher multipolar orders, $\alpha$ notably increases and exhibits a good approximation to the relation $\alpha\approx 2n$ for $n=1,2,3$. In the TM limit, the lifetime scaling converts to
\begin{eqnarray}
\label{eq.TM_scaling}
    t^*\sim \exp(C[\ln(K^{-1}/g)]^2),
\end{eqnarray}
where the constant $C\approx 0.8$ and $g\approx 0.3$ as shown in Fig.~\ref{fig-B0}(c). A similar functional form has been reported in Ref.~\cite{mori2021rigorous} but in the high-frequency regime. We verify that this scaling grows faster than any power law (cf. Sec.~\ref{sec.tm}), indicating a significant suppression of heating in a non-perturbative manner.

\textit{Linearization.---}
Although it is expected that higher multipolar orders may further suppress heating, the perturbative expansion of the effective Hamiltonian is insufficient to explain the scaling of the prethermal lifetime. To address this, we develop a simple theory by linearizing the many-body systems.

Assuming small angular differences between neighboring rotors, $(q_j - q_{j+1}) \text{ mod } 2\pi \ll 1$, we can expand the kicked interaction using the quadratic approximation $\cos (q_j - q_{j+1}) \approx 1 - \frac{1}{2}(q_j - q_{j+1})^2$ in the Hamiltonian Eq.~\ref{eq:Hamiltonian_rotor}. Performing a Fourier transform, we obtain
\begin{equation}
\begin{aligned}
\label{eq.decouplingHamiltonian}
    H(t) = \frac{1}{2} \sum_w\big[\left|P_w\right|^2 \pm F(w)\left|Q_w\right|^2 \sum_{l}\delta(t-l\tau)\big],\ 
\end{aligned}
\end{equation}
where $w := 2\pi I/N$ for an integer $I$, $F(w) := 4K\sin^2(w/2)$, and the Fourier components are defined as $P_w = \sum_{j = 1}^N p_j e^{-iwj}/{\sqrt{N}}$ and $Q_w = \sum_{j = 1}^N q_j e^{-iwj}/{\sqrt{N}}$. The $\pm$ sign follows the RMD sequence. The system now decouples into a set of independent kicked harmonic oscillators labeled by $w$. For each oscillator, we can integrate its discrete time evolution in a two-dimensional phase space over one period $\tau$:
\begin{equation}
\left(\begin{array}{l}
Q_{w} \\
P_{-w}
\end{array}\right)_{t+\tau}=M_0^{\pm}(w)
\left(\begin{array}{l}
Q_{w} \\
P_{-w}
\end{array}\right)_t,
\end{equation}
where $M_0^{\pm}$ is the elementary evolution matrix (see derivations in Sec.~\ref{sec.linearization}):
\begin{equation}
M_0^{\pm} = \begin{pmatrix} 
1 \mp \tau F & \tau \\
\mp F & 1 
\end{pmatrix},
\end{equation} 
where we drop the label $w$ as the following discussion equally applies to all $w$.

\begin{figure}
\centering
\includegraphics[width=0.98\linewidth]{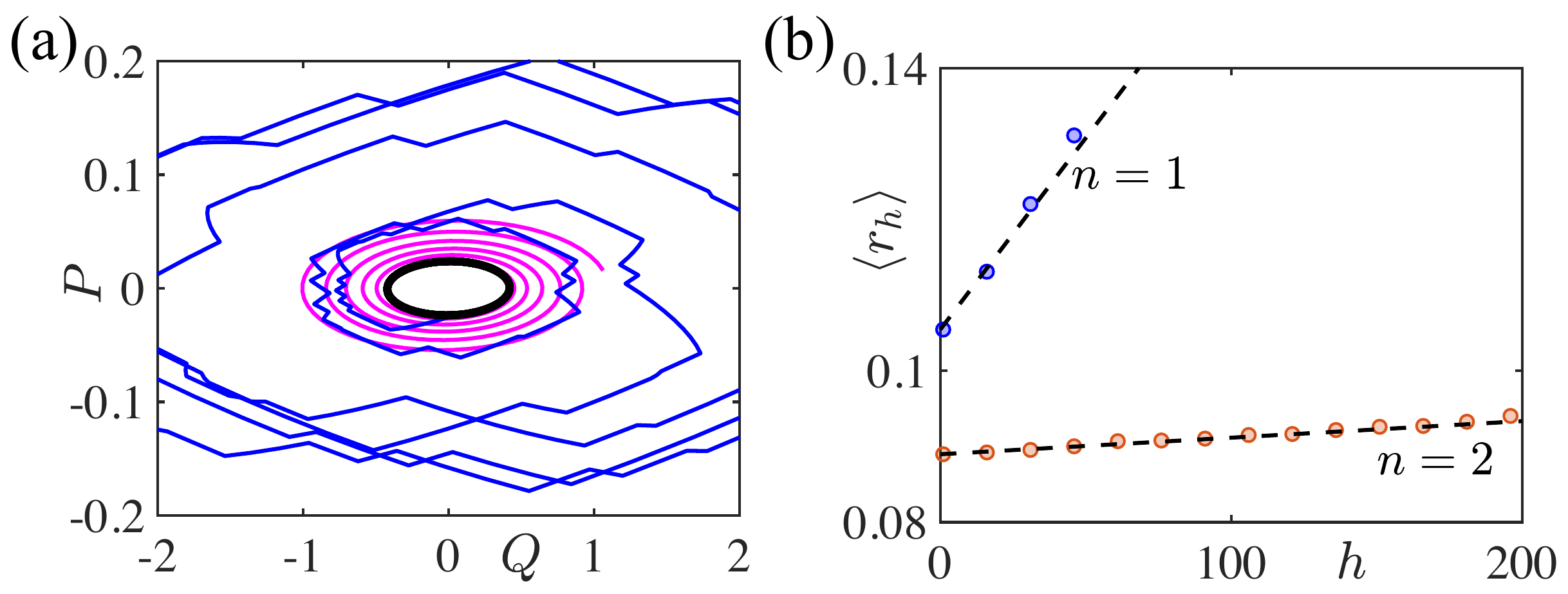}
\caption{(a) Trajectories in phase space. The black orbit is obtained by the area-preserving map $\bar{M}'$. The magenta curve with a constant expansion rate is generated via $\bar{M}_1$. The blue curve is a single realization obtained by stochastically applying the matrix $M^{\pm}_{ 1}$. (b) The averaged radius $\langle r_h \rangle$ matches well with the theoretical prediction (dashed lineas). $F=0.08$ is used in both panels. }
\label{fig-linear}
\end{figure}

Similar to Eq.~\ref{eq:recursive_relation}, higher multipolar evolution matrices can be recursively derived as $M_{n}^{\pm} = M_{n-1}^{\mp}M_{n-1}^{\pm}$ to generate stroboscopic time evolution over the duration $2^n\tau$. Crucially, both $M_n^+$ and $M_n^-$ have the property $\mathrm{det}(M_n^{\pm})=1$, making them area-preserving maps~\cite{kruscha2012biased}. Therefore, when only $M_n^+$ or $M_n^-$ is periodically applied, the system, for weak kicking strength, exhibits non-chaotic dynamics confined to a closed elliptical orbit around its fixed point $(Q,P)=(0,0)$. However, the random concatenation of two slightly different maps $M_n^{\pm}$ generally perturbs these stable trajectories, causing them to deviate indefinitely from their fixed points (Fig.~\ref{fig-linear}(a), blue). By quantifying such deviation, one can estimate the heating rate and its relation to the multipolar order.

To analyze this deviation, we define the averaged evolution matrix as $\bar{M}_n := \frac{1}{2} (M_n^+ + M_n^-)$ and the difference between the two matrices as $D_n := \frac{1}{2}(M_n^+ - M_n^-)$, such that $M_n^{\pm}=\bar{M}_n\pm D_n$. It is worth noting that $\det(\bar{M}_n) = 1+\mathcal{O}((\tau F)^{2n})$ for non-zero $n$, indicating that the averaged map $\bar{M}_n$ does not preserve area in phase space. Instead, the trajectory slowly spirals out with a constant expansion rate scaling as $F^{2n}$ (magenta in Fig.~\ref{fig-linear}(a)). Additionally, the stochastic term $D_n$ possesses eigenvalues that scale as $F^n$ (cf. Sec.~\ref{sec.evalues}), and one would expect it to contribute to a diffusive spiral-out process with a rate also scaling as $F^{2n}$.

To quantify this process, we introduce the normalized map $\bar{M}'={\bar{M}_n}/{\sqrt{\det{\bar{M}_n}}}$, ensuring its area-preserving property with $\det (\bar{M}')=1$, thus generating a closed elliptical orbit (Fig.~\ref{fig-linear}(a), black). The matrix elements of $\bar{M}'$ define the metric of the orbit and determine its conserved area $A(Q, P)$~\footnote{The area is given by the expression 
\begin{align*}
A(Q, P)=\frac{\pi\left[M_{12} P^2-M_{21} Q^2+\left(M_{11}-M_{22}\right) Q P\right]}{\sqrt{1-\left(\frac{M_{11}+M_{22}}{2}\right)^2}}, 
\end{align*}
as a function of $Q$ and $P$ and $M_{ij}$ denotes the matrix elements of $\bar{M}'$~\cite{lichtenberg2013regular}.
}. The radius of the ellipse, defined as $
r_h=\sqrt{{A(Q, P)}/{\pi}}
$, becomes time-dependent when $M_n^{\pm}$ is stochastically applied $h$ times. The expansion rate of the radius can be calculated as ${\Delta r_h}/{r_h}$, where $\Delta r_h = r_{h+1} - r_h$. By averaging over different random realizations and the polar angle of the ellipse, we find that its leading order contribution scales as $F^{2n}$, with a specific expression $\langle {\Delta r_h}/{r_h}\rangle\approx 3 \tau^2 F^{2}/4$ for $n=1$ and $6\tau^4 F^4$ for $n=2$, as detailed in Sec.~\ref{sec.stability_orbits}. Consequently, the averaged growth of the radius at early times can be obtained accordingly.

In Fig.~\ref{fig-linear}(b), we present numerical simulations (circles) of the averaged radius for $n=1$ (blue) and $2$ (orange), which closely match our analytical predictions (dashed lines). As $F$ is proportional to the kicking strength, the expansion rate scales as $K^{2n}$, and its inverse corresponds to the observed prethermal lifetime scaling in Fig.~\ref{fig-B0}.

We note that the strong dependence of the multipolar order $n$ in the prethermal lifetime scaling is remarkably robust, even for initial states that deviate significantly from the linearization regime where $(q_j - q_{j+1}) \text{ mod } 2\pi \ll 1$. Indeed, our numerical results in Fig.~\ref{fig-B0} are obtained using a random distribution of $q_j$ over a wide range $[0,2\pi]$. In Sec.~\ref{sec.momentum_distribution}, we confirm that this phenomenon persists as long as the prethermal regime exhibits a low temperature, leading to a narrow distribution of angular momenta.

\textit{Discussion.---}
We have proposed a mechanism to suppress heating in aperiodically kicked systems by introducing self-similar multipolar structures. This mechanism brings about significant changes in the thermalization pathways and effectively blocks a series of heating channels. As a result, it supports the existence of a long-lived prethermal regime even in the absence of TTS and away from the high-frequency regime.

To demonstrate this mechanism, we have considered classical many-body rotor systems, where we have discovered a characteristic prethermal lifetime scaling of $(1/K)^{2n}$. In the quasi-periodic TM limit, the heating suppression becomes non-perturbative, leading to the scaling given by Eq.~\ref{eq.TM_scaling}, which does not follow an exponential or algebraic form. A similar functional form has been rigorously proven in the high-frequency regime \cite{mori2021rigorous}. However, it remains an interesting open question to justify such a dependence on the kicking strength.

The Hamiltonian Eq.~\ref{eq:Hamiltonian_rotor} can be experimentally realized, e.g., using an array of bosonic Josephson junctions \cite{cataliotti2001josephson,bloch2008many,cao2022interaction}. This opens up possibilities for the experimental exploration of prethermalization with RMD kicks.

\begin{figure}
\centering
\includegraphics[width=0.98\linewidth]{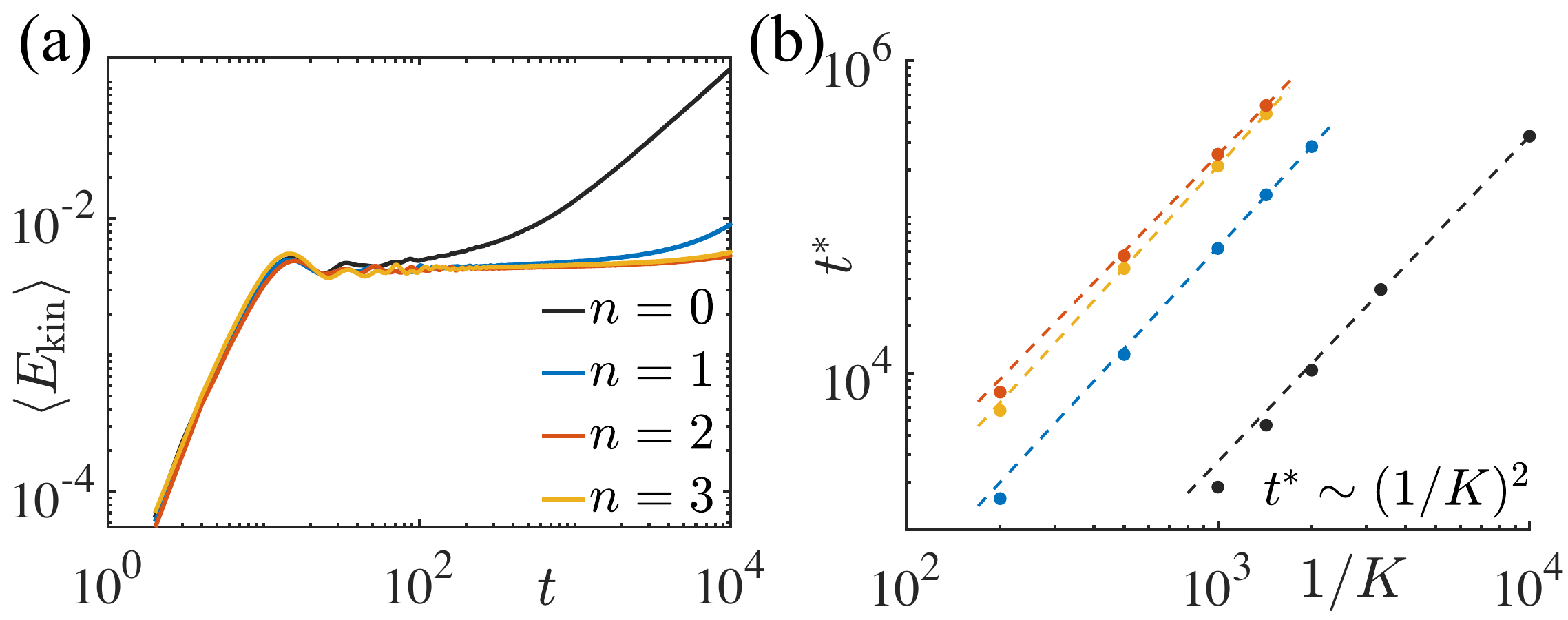}
\caption{(a) Time evolution of the averaged kinetic energy density for different multipolar order $n$ with $K=0.005$ in a log-log scale. (b) Prethermal lifetime $t^*$ as a function of $1/K$ in a log-log scale. 
The system starts from the initial angular momentum $p_j(0)=0$ with $B=0.01$. Other numerical parameters are same as in Fig.~\ref{fig-B0}. Dashed lines correspond to the scaling $K^{-2}$.
 }
\label{fig-B0pt01}
\end{figure}

It is important to note that while heating channels can be suppressed by the RMD sequence, the lifetime scaling is not universal and can strongly depend on the microscopic details of the kicked system. The strong dependence of the multipolar order $n$ in the lifetime scaling may not occur if interaction terms are also present in the static Hamiltonian, such as $H=H_{\mathrm{kin}}+V_{\mathrm{int}}$. The leading order perturbation $\Omega_{n,1}$ involves terms with more than one $V_{\mathrm{int}}$, e.g., $\{V_{\mathrm{int}},\{V_{\mathrm{int}},H_{\mathrm{kin}}\}\}\sim \sum_j\left[\sin(q_j-q_{j+1})-\sin(q_{j-1}-q_j)\right]^2$. These terms are independent of angular momenta and cannot be suppressed even at low prethermal temperatures. Hence, the condition Eq.~\ref{eq:condition} cannot be satisfied in this case.

We implemented a kicked protocol with a modified kick strength $K_l=\pm K+B$ such that the additional static interaction can be efficiently simulated at stroboscopic times~\footnote{Note that simulating the time evolution with a static interaction requires discretizing the continuous equations of motion, which significantly increases the numerical cost for long simulations of the dynamics. Instead, we modify the kick amplitude of the interaction $V$ as $K_l=\pm K+B$, allowing us to generate this interaction approximately at stroboscopic times.}. In Fig.~\ref{fig-B0pt01}(a), we illustrate the results with $B=0.01$. The prethermal plateau is still observed, and the corresponding lifetime is shown in panel (b). It is evident that for $n>0$, heating can still be significantly suppressed. However, its dependence on the kicking strength now follows $t^* \sim (1/K)^2$ regardless of the multipolar order. A similar linearization analysis can be performed, and the expansion rate for each decoupled oscillator is $K^2$, as detailed in Sec.~\ref{sec.evalues}. Identifying a general mechanism for further suppressing heating with a better scaling remains an intriguing open question.

Finally, we note that the perturbative expansion predicting the  suppression of heating also applies to RMD kicked {\it quantum} systems. A systematic study of quantum thermalization in kicked systems and its relation to their classical counterparts is
an intriguing subject for future study.

\textit{Acknowledgements.---}
We thank Marin Bukov, Johannes Knolle, Holger Kantz, Roland Ketzmerick, Paul Schindler and Yujie Liu for many useful discussions. This work is in part supported by the Deutsche Forschungsgemeinschaft under cluster
of excellence ct.qmat (EXC 2147, project-id 390858490).

\bibliography{Reference}

\begin{thebibliography}{74}%
\makeatletter
\providecommand \@ifxundefined [1]{%
 \@ifx{#1\undefined}
}%
\providecommand \@ifnum [1]{%
 \ifnum #1\expandafter \@firstoftwo
 \else \expandafter \@secondoftwo
 \fi
}%
\providecommand \@ifx [1]{%
 \ifx #1\expandafter \@firstoftwo
 \else \expandafter \@secondoftwo
 \fi
}%
\providecommand \natexlab [1]{#1}%
\providecommand \enquote  [1]{``#1''}%
\providecommand \bibnamefont  [1]{#1}%
\providecommand \bibfnamefont [1]{#1}%
\providecommand \citenamefont [1]{#1}%
\providecommand \href@noop [0]{\@secondoftwo}%
\providecommand \href [0]{\begingroup \@sanitize@url \@href}%
\providecommand \@href[1]{\@@startlink{#1}\@@href}%
\providecommand \@@href[1]{\endgroup#1\@@endlink}%
\providecommand \@sanitize@url [0]{\catcode `\\12\catcode `\$12\catcode
  `\&12\catcode `\#12\catcode `\^12\catcode `\_12\catcode `\%12\relax}%
\providecommand \@@startlink[1]{}%
\providecommand \@@endlink[0]{}%
\providecommand \url  [0]{\begingroup\@sanitize@url \@url }%
\providecommand \@url [1]{\endgroup\@href {#1}{\urlprefix }}%
\providecommand \urlprefix  [0]{URL }%
\providecommand \Eprint [0]{\href }%
\providecommand \doibase [0]{https://doi.org/}%
\providecommand \selectlanguage [0]{\@gobble}%
\providecommand \bibinfo  [0]{\@secondoftwo}%
\providecommand \bibfield  [0]{\@secondoftwo}%
\providecommand \translation [1]{[#1]}%
\providecommand \BibitemOpen [0]{}%
\providecommand \bibitemStop [0]{}%
\providecommand \bibitemNoStop [0]{.\EOS\space}%
\providecommand \EOS [0]{\spacefactor3000\relax}%
\providecommand \BibitemShut  [1]{\csname bibitem#1\endcsname}%
\let\auto@bib@innerbib\@empty
\bibitem [{\citenamefont {Khemani}\ \emph {et~al.}(2019)\citenamefont
  {Khemani}, \citenamefont {Moessner},\ and\ \citenamefont
  {Sondhi}}]{khemani2019brief}%
  \BibitemOpen
  \bibfield  {author} {\bibinfo {author} {\bibfnamefont {V.}~\bibnamefont
  {Khemani}}, \bibinfo {author} {\bibfnamefont {R.}~\bibnamefont {Moessner}},\
  and\ \bibinfo {author} {\bibfnamefont {S.}~\bibnamefont {Sondhi}},\
  }\bibfield  {title} {\bibinfo {title} {A brief history of time crystals},\
  }\href@noop {} {\bibfield  {journal} {\bibinfo  {journal} {arXiv preprint
  arXiv:1910.10745}\ } (\bibinfo {year} {2019})}\BibitemShut {NoStop}%
\bibitem [{\citenamefont {Zaletel}\ \emph {et~al.}(2023)\citenamefont
  {Zaletel}, \citenamefont {Lukin}, \citenamefont {Monroe}, \citenamefont
  {Nayak}, \citenamefont {Wilczek},\ and\ \citenamefont
  {Yao}}]{zaletel2023colloquium}%
  \BibitemOpen
  \bibfield  {author} {\bibinfo {author} {\bibfnamefont {M.~P.}\ \bibnamefont
  {Zaletel}}, \bibinfo {author} {\bibfnamefont {M.}~\bibnamefont {Lukin}},
  \bibinfo {author} {\bibfnamefont {C.}~\bibnamefont {Monroe}}, \bibinfo
  {author} {\bibfnamefont {C.}~\bibnamefont {Nayak}}, \bibinfo {author}
  {\bibfnamefont {F.}~\bibnamefont {Wilczek}},\ and\ \bibinfo {author}
  {\bibfnamefont {N.~Y.}\ \bibnamefont {Yao}},\ }\bibfield  {title} {\bibinfo
  {title} {Colloquium: Quantum and classical discrete time crystals},\
  }\href@noop {} {\bibfield  {journal} {\bibinfo  {journal} {arXiv preprint
  arXiv:2305.08904}\ } (\bibinfo {year} {2023})}\BibitemShut {NoStop}%
\bibitem [{\citenamefont {D’Alessio}\ and\ \citenamefont
  {Rigol}(2014)}]{d2014long}%
  \BibitemOpen
  \bibfield  {author} {\bibinfo {author} {\bibfnamefont {L.}~\bibnamefont
  {D’Alessio}}\ and\ \bibinfo {author} {\bibfnamefont {M.}~\bibnamefont
  {Rigol}},\ }\bibfield  {title} {\bibinfo {title} {Long-time behavior of
  isolated periodically driven interacting lattice systems},\ }\href@noop {}
  {\bibfield  {journal} {\bibinfo  {journal} {Physical Review X}\ }\textbf
  {\bibinfo {volume} {4}},\ \bibinfo {pages} {041048} (\bibinfo {year}
  {2014})}\BibitemShut {NoStop}%
\bibitem [{\citenamefont {Lazarides}\ \emph {et~al.}(2014)\citenamefont
  {Lazarides}, \citenamefont {Das},\ and\ \citenamefont
  {Moessner}}]{lazarides2014equilibrium}%
  \BibitemOpen
  \bibfield  {author} {\bibinfo {author} {\bibfnamefont {A.}~\bibnamefont
  {Lazarides}}, \bibinfo {author} {\bibfnamefont {A.}~\bibnamefont {Das}},\
  and\ \bibinfo {author} {\bibfnamefont {R.}~\bibnamefont {Moessner}},\
  }\bibfield  {title} {\bibinfo {title} {Equilibrium states of generic quantum
  systems subject to periodic driving},\ }\href@noop {} {\bibfield  {journal}
  {\bibinfo  {journal} {Physical Review E}\ }\textbf {\bibinfo {volume} {90}},\
  \bibinfo {pages} {012110} (\bibinfo {year} {2014})}\BibitemShut {NoStop}%
\bibitem [{\citenamefont {Bukov}\ \emph {et~al.}(2015)\citenamefont {Bukov},
  \citenamefont {D'Alessio},\ and\ \citenamefont
  {Polkovnikov}}]{bukov2015universal}%
  \BibitemOpen
  \bibfield  {author} {\bibinfo {author} {\bibfnamefont {M.}~\bibnamefont
  {Bukov}}, \bibinfo {author} {\bibfnamefont {L.}~\bibnamefont {D'Alessio}},\
  and\ \bibinfo {author} {\bibfnamefont {A.}~\bibnamefont {Polkovnikov}},\
  }\bibfield  {title} {\bibinfo {title} {Universal high-frequency behavior of
  periodically driven systems: from dynamical stabilization to floquet
  engineering},\ }\href@noop {} {\bibfield  {journal} {\bibinfo  {journal}
  {Advances in Physics}\ }\textbf {\bibinfo {volume} {64}},\ \bibinfo {pages}
  {139} (\bibinfo {year} {2015})}\BibitemShut {NoStop}%
\bibitem [{\citenamefont {Abanin}\ \emph {et~al.}(2015)\citenamefont {Abanin},
  \citenamefont {De~Roeck},\ and\ \citenamefont
  {Huveneers}}]{abanin2015exponentially}%
  \BibitemOpen
  \bibfield  {author} {\bibinfo {author} {\bibfnamefont {D.~A.}\ \bibnamefont
  {Abanin}}, \bibinfo {author} {\bibfnamefont {W.}~\bibnamefont {De~Roeck}},\
  and\ \bibinfo {author} {\bibfnamefont {F.}~\bibnamefont {Huveneers}},\
  }\bibfield  {title} {\bibinfo {title} {Exponentially slow heating in
  periodically driven many-body systems},\ }\href@noop {} {\bibfield  {journal}
  {\bibinfo  {journal} {Physical review letters}\ }\textbf {\bibinfo {volume}
  {115}},\ \bibinfo {pages} {256803} (\bibinfo {year} {2015})}\BibitemShut
  {NoStop}%
\bibitem [{\citenamefont {Kuwahara}\ \emph {et~al.}(2016)\citenamefont
  {Kuwahara}, \citenamefont {Mori},\ and\ \citenamefont
  {Saito}}]{kuwahara2016floquet}%
  \BibitemOpen
  \bibfield  {author} {\bibinfo {author} {\bibfnamefont {T.}~\bibnamefont
  {Kuwahara}}, \bibinfo {author} {\bibfnamefont {T.}~\bibnamefont {Mori}},\
  and\ \bibinfo {author} {\bibfnamefont {K.}~\bibnamefont {Saito}},\ }\bibfield
   {title} {\bibinfo {title} {Floquet--magnus theory and generic transient
  dynamics in periodically driven many-body quantum systems},\ }\href@noop {}
  {\bibfield  {journal} {\bibinfo  {journal} {Annals of Physics}\ }\textbf
  {\bibinfo {volume} {367}},\ \bibinfo {pages} {96} (\bibinfo {year}
  {2016})}\BibitemShut {NoStop}%
\bibitem [{\citenamefont {Weidinger}\ and\ \citenamefont
  {Knap}(2017)}]{weidinger2017floquet}%
  \BibitemOpen
  \bibfield  {author} {\bibinfo {author} {\bibfnamefont {S.~A.}\ \bibnamefont
  {Weidinger}}\ and\ \bibinfo {author} {\bibfnamefont {M.}~\bibnamefont
  {Knap}},\ }\bibfield  {title} {\bibinfo {title} {Floquet prethermalization
  and regimes of heating in a periodically driven, interacting quantum
  system},\ }\href@noop {} {\bibfield  {journal} {\bibinfo  {journal}
  {Scientific reports}\ }\textbf {\bibinfo {volume} {7}},\ \bibinfo {pages} {1}
  (\bibinfo {year} {2017})}\BibitemShut {NoStop}%
\bibitem [{\citenamefont {Sieberer}\ \emph {et~al.}(2019)\citenamefont
  {Sieberer}, \citenamefont {Olsacher}, \citenamefont {Elben}, \citenamefont
  {Heyl}, \citenamefont {Hauke}, \citenamefont {Haake},\ and\ \citenamefont
  {Zoller}}]{sieberer2019digital}%
  \BibitemOpen
  \bibfield  {author} {\bibinfo {author} {\bibfnamefont {L.~M.}\ \bibnamefont
  {Sieberer}}, \bibinfo {author} {\bibfnamefont {T.}~\bibnamefont {Olsacher}},
  \bibinfo {author} {\bibfnamefont {A.}~\bibnamefont {Elben}}, \bibinfo
  {author} {\bibfnamefont {M.}~\bibnamefont {Heyl}}, \bibinfo {author}
  {\bibfnamefont {P.}~\bibnamefont {Hauke}}, \bibinfo {author} {\bibfnamefont
  {F.}~\bibnamefont {Haake}},\ and\ \bibinfo {author} {\bibfnamefont
  {P.}~\bibnamefont {Zoller}},\ }\bibfield  {title} {\bibinfo {title} {Digital
  quantum simulation, trotter errors, and quantum chaos of the kicked top},\
  }\href@noop {} {\bibfield  {journal} {\bibinfo  {journal} {npj Quantum
  Information}\ }\textbf {\bibinfo {volume} {5}},\ \bibinfo {pages} {78}
  (\bibinfo {year} {2019})}\BibitemShut {NoStop}%
\bibitem [{\citenamefont {Rubio-Abadal}\ \emph {et~al.}(2020)\citenamefont
  {Rubio-Abadal}, \citenamefont {Ippoliti}, \citenamefont {Hollerith},
  \citenamefont {Wei}, \citenamefont {Rui}, \citenamefont {Sondhi},
  \citenamefont {Khemani}, \citenamefont {Gross},\ and\ \citenamefont
  {Bloch}}]{rubio2020floquet}%
  \BibitemOpen
  \bibfield  {author} {\bibinfo {author} {\bibfnamefont {A.}~\bibnamefont
  {Rubio-Abadal}}, \bibinfo {author} {\bibfnamefont {M.}~\bibnamefont
  {Ippoliti}}, \bibinfo {author} {\bibfnamefont {S.}~\bibnamefont {Hollerith}},
  \bibinfo {author} {\bibfnamefont {D.}~\bibnamefont {Wei}}, \bibinfo {author}
  {\bibfnamefont {J.}~\bibnamefont {Rui}}, \bibinfo {author} {\bibfnamefont
  {S.}~\bibnamefont {Sondhi}}, \bibinfo {author} {\bibfnamefont
  {V.}~\bibnamefont {Khemani}}, \bibinfo {author} {\bibfnamefont
  {C.}~\bibnamefont {Gross}},\ and\ \bibinfo {author} {\bibfnamefont
  {I.}~\bibnamefont {Bloch}},\ }\bibfield  {title} {\bibinfo {title} {Floquet
  prethermalization in a bose-hubbard system},\ }\href@noop {} {\bibfield
  {journal} {\bibinfo  {journal} {Physical Review X}\ }\textbf {\bibinfo
  {volume} {10}},\ \bibinfo {pages} {021044} (\bibinfo {year}
  {2020})}\BibitemShut {NoStop}%
\bibitem [{\citenamefont {Hodson}\ and\ \citenamefont
  {Jarzynski}(2021)}]{hodson2021energy}%
  \BibitemOpen
  \bibfield  {author} {\bibinfo {author} {\bibfnamefont {W.}~\bibnamefont
  {Hodson}}\ and\ \bibinfo {author} {\bibfnamefont {C.}~\bibnamefont
  {Jarzynski}},\ }\bibfield  {title} {\bibinfo {title} {Energy diffusion and
  absorption in chaotic systems with rapid periodic driving},\ }\href@noop {}
  {\bibfield  {journal} {\bibinfo  {journal} {Physical Review Research}\
  }\textbf {\bibinfo {volume} {3}},\ \bibinfo {pages} {013219} (\bibinfo {year}
  {2021})}\BibitemShut {NoStop}%
\bibitem [{\citenamefont {McRoberts}\ \emph {et~al.}(2022)\citenamefont
  {McRoberts}, \citenamefont {Zhao}, \citenamefont {Moessner},\ and\
  \citenamefont {Bukov}}]{mcroberts2022prethermalization}%
  \BibitemOpen
  \bibfield  {author} {\bibinfo {author} {\bibfnamefont {A.~J.}\ \bibnamefont
  {McRoberts}}, \bibinfo {author} {\bibfnamefont {H.}~\bibnamefont {Zhao}},
  \bibinfo {author} {\bibfnamefont {R.}~\bibnamefont {Moessner}},\ and\
  \bibinfo {author} {\bibfnamefont {M.}~\bibnamefont {Bukov}},\ }\bibfield
  {title} {\bibinfo {title} {'prethermalization'in conservative nonsymplectic
  periodically driven spin systems},\ }\href@noop {} {\bibfield  {journal}
  {\bibinfo  {journal} {arXiv preprint arXiv:2208.09005}\ } (\bibinfo {year}
  {2022})}\BibitemShut {NoStop}%
\bibitem [{\citenamefont {Beatrez}\ \emph {et~al.}(2023)\citenamefont
  {Beatrez}, \citenamefont {Fleckenstein}, \citenamefont {Pillai},
  \citenamefont {de~Leon~Sanchez}, \citenamefont {Akkiraju}, \citenamefont
  {Diaz~Alcala}, \citenamefont {Conti}, \citenamefont {Reshetikhin},
  \citenamefont {Druga}, \citenamefont {Bukov} \emph
  {et~al.}}]{beatrez2023critical}%
  \BibitemOpen
  \bibfield  {author} {\bibinfo {author} {\bibfnamefont {W.}~\bibnamefont
  {Beatrez}}, \bibinfo {author} {\bibfnamefont {C.}~\bibnamefont
  {Fleckenstein}}, \bibinfo {author} {\bibfnamefont {A.}~\bibnamefont
  {Pillai}}, \bibinfo {author} {\bibfnamefont {E.}~\bibnamefont
  {de~Leon~Sanchez}}, \bibinfo {author} {\bibfnamefont {A.}~\bibnamefont
  {Akkiraju}}, \bibinfo {author} {\bibfnamefont {J.}~\bibnamefont
  {Diaz~Alcala}}, \bibinfo {author} {\bibfnamefont {S.}~\bibnamefont {Conti}},
  \bibinfo {author} {\bibfnamefont {P.}~\bibnamefont {Reshetikhin}}, \bibinfo
  {author} {\bibfnamefont {E.}~\bibnamefont {Druga}}, \bibinfo {author}
  {\bibfnamefont {M.}~\bibnamefont {Bukov}}, \emph {et~al.},\ }\bibfield
  {title} {\bibinfo {title} {Critical prethermal discrete time crystal created
  by two-frequency driving},\ }\href@noop {} {\bibfield  {journal} {\bibinfo
  {journal} {Nature Physics}\ ,\ \bibinfo {pages} {1}} (\bibinfo {year}
  {2023})}\BibitemShut {NoStop}%
\bibitem [{\citenamefont {Ho}\ \emph {et~al.}(2023)\citenamefont {Ho},
  \citenamefont {Mori}, \citenamefont {Abanin},\ and\ \citenamefont
  {Dalla~Torre}}]{ho2023quantum}%
  \BibitemOpen
  \bibfield  {author} {\bibinfo {author} {\bibfnamefont {W.~W.}\ \bibnamefont
  {Ho}}, \bibinfo {author} {\bibfnamefont {T.}~\bibnamefont {Mori}}, \bibinfo
  {author} {\bibfnamefont {D.~A.}\ \bibnamefont {Abanin}},\ and\ \bibinfo
  {author} {\bibfnamefont {E.~G.}\ \bibnamefont {Dalla~Torre}},\ }\bibfield
  {title} {\bibinfo {title} {Quantum and classical floquet prethermalization},\
  }\href@noop {} {\bibfield  {journal} {\bibinfo  {journal} {Annals of
  Physics}\ ,\ \bibinfo {pages} {169297}} (\bibinfo {year} {2023})}\BibitemShut
  {NoStop}%
\bibitem [{\citenamefont {Vajna}\ \emph {et~al.}(2018)\citenamefont {Vajna},
  \citenamefont {Klobas}, \citenamefont {Prosen},\ and\ \citenamefont
  {Polkovnikov}}]{vajna2018replica}%
  \BibitemOpen
  \bibfield  {author} {\bibinfo {author} {\bibfnamefont {S.}~\bibnamefont
  {Vajna}}, \bibinfo {author} {\bibfnamefont {K.}~\bibnamefont {Klobas}},
  \bibinfo {author} {\bibfnamefont {T.}~\bibnamefont {Prosen}},\ and\ \bibinfo
  {author} {\bibfnamefont {A.}~\bibnamefont {Polkovnikov}},\ }\bibfield
  {title} {\bibinfo {title} {Replica resummation of the
  baker-campbell-hausdorff series},\ }\href@noop {} {\bibfield  {journal}
  {\bibinfo  {journal} {Physical review letters}\ }\textbf {\bibinfo {volume}
  {120}},\ \bibinfo {pages} {200607} (\bibinfo {year} {2018})}\BibitemShut
  {NoStop}%
\bibitem [{\citenamefont {Fleckenstein}\ and\ \citenamefont
  {Bukov}(2021)}]{fleckenstein2021thermalization}%
  \BibitemOpen
  \bibfield  {author} {\bibinfo {author} {\bibfnamefont {C.}~\bibnamefont
  {Fleckenstein}}\ and\ \bibinfo {author} {\bibfnamefont {M.}~\bibnamefont
  {Bukov}},\ }\bibfield  {title} {\bibinfo {title} {Thermalization and
  prethermalization in periodically kicked quantum spin chains},\ }\href@noop
  {} {\bibfield  {journal} {\bibinfo  {journal} {Physical Review B}\ }\textbf
  {\bibinfo {volume} {103}},\ \bibinfo {pages} {144307} (\bibinfo {year}
  {2021})}\BibitemShut {NoStop}%
\bibitem [{\citenamefont {Mori}(2022)}]{mori2022heating}%
  \BibitemOpen
  \bibfield  {author} {\bibinfo {author} {\bibfnamefont {T.}~\bibnamefont
  {Mori}},\ }\bibfield  {title} {\bibinfo {title} {Heating rates under fast
  periodic driving beyond linear response},\ }\href@noop {} {\bibfield
  {journal} {\bibinfo  {journal} {Physical Review Letters}\ }\textbf {\bibinfo
  {volume} {128}},\ \bibinfo {pages} {050604} (\bibinfo {year}
  {2022})}\BibitemShut {NoStop}%
\bibitem [{\citenamefont {Mori}(2018)}]{mori2018floquet}%
  \BibitemOpen
  \bibfield  {author} {\bibinfo {author} {\bibfnamefont {T.}~\bibnamefont
  {Mori}},\ }\bibfield  {title} {\bibinfo {title} {Floquet prethermalization in
  periodically driven classical spin systems},\ }\href@noop {} {\bibfield
  {journal} {\bibinfo  {journal} {Physical Review B}\ }\textbf {\bibinfo
  {volume} {98}},\ \bibinfo {pages} {104303} (\bibinfo {year}
  {2018})}\BibitemShut {NoStop}%
\bibitem [{\citenamefont {Haldar}\ \emph {et~al.}(2018)\citenamefont {Haldar},
  \citenamefont {Moessner},\ and\ \citenamefont {Das}}]{haldar2018onset}%
  \BibitemOpen
  \bibfield  {author} {\bibinfo {author} {\bibfnamefont {A.}~\bibnamefont
  {Haldar}}, \bibinfo {author} {\bibfnamefont {R.}~\bibnamefont {Moessner}},\
  and\ \bibinfo {author} {\bibfnamefont {A.}~\bibnamefont {Das}},\ }\bibfield
  {title} {\bibinfo {title} {Onset of floquet thermalization},\ }\href@noop {}
  {\bibfield  {journal} {\bibinfo  {journal} {Physical Review B}\ }\textbf
  {\bibinfo {volume} {97}},\ \bibinfo {pages} {245122} (\bibinfo {year}
  {2018})}\BibitemShut {NoStop}%
\bibitem [{\citenamefont {Howell}\ \emph {et~al.}(2019)\citenamefont {Howell},
  \citenamefont {Weinberg}, \citenamefont {Sels}, \citenamefont {Polkovnikov},\
  and\ \citenamefont {Bukov}}]{howell2019asymptotic}%
  \BibitemOpen
  \bibfield  {author} {\bibinfo {author} {\bibfnamefont {O.}~\bibnamefont
  {Howell}}, \bibinfo {author} {\bibfnamefont {P.}~\bibnamefont {Weinberg}},
  \bibinfo {author} {\bibfnamefont {D.}~\bibnamefont {Sels}}, \bibinfo {author}
  {\bibfnamefont {A.}~\bibnamefont {Polkovnikov}},\ and\ \bibinfo {author}
  {\bibfnamefont {M.}~\bibnamefont {Bukov}},\ }\bibfield  {title} {\bibinfo
  {title} {Asymptotic prethermalization in periodically driven classical spin
  chains},\ }\href@noop {} {\bibfield  {journal} {\bibinfo  {journal} {Physical
  review letters}\ }\textbf {\bibinfo {volume} {122}},\ \bibinfo {pages}
  {010602} (\bibinfo {year} {2019})}\BibitemShut {NoStop}%
\bibitem [{\citenamefont {Pizzi}\ \emph {et~al.}(2021)\citenamefont {Pizzi},
  \citenamefont {Nunnenkamp},\ and\ \citenamefont
  {Knolle}}]{pizzi2021classical}%
  \BibitemOpen
  \bibfield  {author} {\bibinfo {author} {\bibfnamefont {A.}~\bibnamefont
  {Pizzi}}, \bibinfo {author} {\bibfnamefont {A.}~\bibnamefont {Nunnenkamp}},\
  and\ \bibinfo {author} {\bibfnamefont {J.}~\bibnamefont {Knolle}},\
  }\bibfield  {title} {\bibinfo {title} {Classical prethermal phases of
  matter},\ }\href@noop {} {\bibfield  {journal} {\bibinfo  {journal} {Physical
  Review Letters}\ }\textbf {\bibinfo {volume} {127}},\ \bibinfo {pages}
  {140602} (\bibinfo {year} {2021})}\BibitemShut {NoStop}%
\bibitem [{\citenamefont {Ikeda}\ and\ \citenamefont
  {Polkovnikov}(2021)}]{ikeda2021fermi}%
  \BibitemOpen
  \bibfield  {author} {\bibinfo {author} {\bibfnamefont {T.~N.}\ \bibnamefont
  {Ikeda}}\ and\ \bibinfo {author} {\bibfnamefont {A.}~\bibnamefont
  {Polkovnikov}},\ }\bibfield  {title} {\bibinfo {title} {Fermi's golden rule
  for heating in strongly driven floquet systems},\ }\href@noop {} {\bibfield
  {journal} {\bibinfo  {journal} {Physical Review B}\ }\textbf {\bibinfo
  {volume} {104}},\ \bibinfo {pages} {134308} (\bibinfo {year}
  {2021})}\BibitemShut {NoStop}%
\bibitem [{\citenamefont {Ye}\ \emph {et~al.}(2021)\citenamefont {Ye},
  \citenamefont {Machado},\ and\ \citenamefont {Yao}}]{ye2021floquet}%
  \BibitemOpen
  \bibfield  {author} {\bibinfo {author} {\bibfnamefont {B.}~\bibnamefont
  {Ye}}, \bibinfo {author} {\bibfnamefont {F.}~\bibnamefont {Machado}},\ and\
  \bibinfo {author} {\bibfnamefont {N.~Y.}\ \bibnamefont {Yao}},\ }\bibfield
  {title} {\bibinfo {title} {Floquet phases of matter via classical
  prethermalization},\ }\href@noop {} {\bibfield  {journal} {\bibinfo
  {journal} {Physical Review Letters}\ }\textbf {\bibinfo {volume} {127}},\
  \bibinfo {pages} {140603} (\bibinfo {year} {2021})}\BibitemShut {NoStop}%
\bibitem [{\citenamefont {Jin}\ \emph {et~al.}(2023)\citenamefont {Jin},
  \citenamefont {Knolle},\ and\ \citenamefont {Knap}}]{jin2023fractionalized}%
  \BibitemOpen
  \bibfield  {author} {\bibinfo {author} {\bibfnamefont {H.-K.}\ \bibnamefont
  {Jin}}, \bibinfo {author} {\bibfnamefont {J.}~\bibnamefont {Knolle}},\ and\
  \bibinfo {author} {\bibfnamefont {M.}~\bibnamefont {Knap}},\ }\bibfield
  {title} {\bibinfo {title} {Fractionalized prethermalization in a driven
  quantum spin liquid},\ }\href@noop {} {\bibfield  {journal} {\bibinfo
  {journal} {Physical Review Letters}\ }\textbf {\bibinfo {volume} {130}},\
  \bibinfo {pages} {226701} (\bibinfo {year} {2023})}\BibitemShut {NoStop}%
\bibitem [{\citenamefont {Lysne}\ \emph {et~al.}(2020)\citenamefont {Lysne},
  \citenamefont {Kuper}, \citenamefont {Poggi}, \citenamefont {Deutsch},\ and\
  \citenamefont {Jessen}}]{lysne2020small}%
  \BibitemOpen
  \bibfield  {author} {\bibinfo {author} {\bibfnamefont {N.~K.}\ \bibnamefont
  {Lysne}}, \bibinfo {author} {\bibfnamefont {K.~W.}\ \bibnamefont {Kuper}},
  \bibinfo {author} {\bibfnamefont {P.~M.}\ \bibnamefont {Poggi}}, \bibinfo
  {author} {\bibfnamefont {I.~H.}\ \bibnamefont {Deutsch}},\ and\ \bibinfo
  {author} {\bibfnamefont {P.~S.}\ \bibnamefont {Jessen}},\ }\bibfield  {title}
  {\bibinfo {title} {Small, highly accurate quantum processor for
  intermediate-depth quantum simulations},\ }\href@noop {} {\bibfield
  {journal} {\bibinfo  {journal} {Physical review letters}\ }\textbf {\bibinfo
  {volume} {124}},\ \bibinfo {pages} {230501} (\bibinfo {year}
  {2020})}\BibitemShut {NoStop}%
\bibitem [{\citenamefont {Joshi}\ \emph {et~al.}(2022)\citenamefont {Joshi},
  \citenamefont {Elben}, \citenamefont {Vikram}, \citenamefont {Vermersch},
  \citenamefont {Galitski},\ and\ \citenamefont {Zoller}}]{joshi2022probing}%
  \BibitemOpen
  \bibfield  {author} {\bibinfo {author} {\bibfnamefont {L.~K.}\ \bibnamefont
  {Joshi}}, \bibinfo {author} {\bibfnamefont {A.}~\bibnamefont {Elben}},
  \bibinfo {author} {\bibfnamefont {A.}~\bibnamefont {Vikram}}, \bibinfo
  {author} {\bibfnamefont {B.}~\bibnamefont {Vermersch}}, \bibinfo {author}
  {\bibfnamefont {V.}~\bibnamefont {Galitski}},\ and\ \bibinfo {author}
  {\bibfnamefont {P.}~\bibnamefont {Zoller}},\ }\bibfield  {title} {\bibinfo
  {title} {Probing many-body quantum chaos with quantum simulators},\
  }\href@noop {} {\bibfield  {journal} {\bibinfo  {journal} {Physical Review
  X}\ }\textbf {\bibinfo {volume} {12}},\ \bibinfo {pages} {011018} (\bibinfo
  {year} {2022})}\BibitemShut {NoStop}%
\bibitem [{\citenamefont {Chirikov}(1971)}]{chirikov1971research}%
  \BibitemOpen
  \bibfield  {author} {\bibinfo {author} {\bibfnamefont {B.~V.}\ \bibnamefont
  {Chirikov}},\ }\href@noop {} {\emph {\bibinfo {title} {Research concerning
  the theory of non-linear resonance and stochasticity}}},\ \bibinfo {type}
  {Tech. Rep.}\ (\bibinfo  {institution} {CM-P00100691},\ \bibinfo {year}
  {1971})\BibitemShut {NoStop}%
\bibitem [{\citenamefont {Haake}\ \emph {et~al.}(1987)\citenamefont {Haake},
  \citenamefont {Ku{\'s}},\ and\ \citenamefont {Scharf}}]{haake1987classical}%
  \BibitemOpen
  \bibfield  {author} {\bibinfo {author} {\bibfnamefont {F.}~\bibnamefont
  {Haake}}, \bibinfo {author} {\bibfnamefont {M.}~\bibnamefont {Ku{\'s}}},\
  and\ \bibinfo {author} {\bibfnamefont {R.}~\bibnamefont {Scharf}},\
  }\bibfield  {title} {\bibinfo {title} {Classical and quantum chaos for a
  kicked top},\ }\href@noop {} {\bibfield  {journal} {\bibinfo  {journal}
  {Zeitschrift f{\"u}r Physik B Condensed Matter}\ }\textbf {\bibinfo {volume}
  {65}},\ \bibinfo {pages} {381} (\bibinfo {year} {1987})}\BibitemShut
  {NoStop}%
\bibitem [{\citenamefont {Kaneko}\ and\ \citenamefont
  {Konishi}(1989)}]{kaneko1989diffusion}%
  \BibitemOpen
  \bibfield  {author} {\bibinfo {author} {\bibfnamefont {K.}~\bibnamefont
  {Kaneko}}\ and\ \bibinfo {author} {\bibfnamefont {T.}~\bibnamefont
  {Konishi}},\ }\bibfield  {title} {\bibinfo {title} {Diffusion in hamiltonian
  dynamical systems with many degrees of freedom},\ }\href@noop {} {\bibfield
  {journal} {\bibinfo  {journal} {Physical Review A}\ }\textbf {\bibinfo
  {volume} {40}},\ \bibinfo {pages} {6130} (\bibinfo {year}
  {1989})}\BibitemShut {NoStop}%
\bibitem [{\citenamefont {Konishi}\ and\ \citenamefont
  {Kaneko}(1990)}]{konishi1990diffusion}%
  \BibitemOpen
  \bibfield  {author} {\bibinfo {author} {\bibfnamefont {T.}~\bibnamefont
  {Konishi}}\ and\ \bibinfo {author} {\bibfnamefont {K.}~\bibnamefont
  {Kaneko}},\ }\bibfield  {title} {\bibinfo {title} {Diffusion in hamiltonian
  chaos and its size dependence},\ }\href@noop {} {\bibfield  {journal}
  {\bibinfo  {journal} {Journal of Physics A: Mathematical and General}\
  }\textbf {\bibinfo {volume} {23}},\ \bibinfo {pages} {L715} (\bibinfo {year}
  {1990})}\BibitemShut {NoStop}%
\bibitem [{\citenamefont {Falcioni}\ \emph {et~al.}(1991)\citenamefont
  {Falcioni}, \citenamefont {Marconi},\ and\ \citenamefont
  {Vulpiani}}]{falcioni1991ergodic}%
  \BibitemOpen
  \bibfield  {author} {\bibinfo {author} {\bibfnamefont {M.}~\bibnamefont
  {Falcioni}}, \bibinfo {author} {\bibfnamefont {U.~M.~B.}\ \bibnamefont
  {Marconi}},\ and\ \bibinfo {author} {\bibfnamefont {A.}~\bibnamefont
  {Vulpiani}},\ }\bibfield  {title} {\bibinfo {title} {Ergodic properties of
  high-dimensional symplectic maps},\ }\href@noop {} {\bibfield  {journal}
  {\bibinfo  {journal} {Physical Review A}\ }\textbf {\bibinfo {volume} {44}},\
  \bibinfo {pages} {2263} (\bibinfo {year} {1991})}\BibitemShut {NoStop}%
\bibitem [{\citenamefont {Mulansky}\ \emph {et~al.}(2011)\citenamefont
  {Mulansky}, \citenamefont {Ahnert}, \citenamefont {Pikovsky},\ and\
  \citenamefont {Shepelyansky}}]{mulansky2011strong}%
  \BibitemOpen
  \bibfield  {author} {\bibinfo {author} {\bibfnamefont {M.}~\bibnamefont
  {Mulansky}}, \bibinfo {author} {\bibfnamefont {K.}~\bibnamefont {Ahnert}},
  \bibinfo {author} {\bibfnamefont {A.}~\bibnamefont {Pikovsky}},\ and\
  \bibinfo {author} {\bibfnamefont {D.}~\bibnamefont {Shepelyansky}},\
  }\bibfield  {title} {\bibinfo {title} {Strong and weak chaos in weakly
  nonintegrable many-body hamiltonian systems},\ }\href@noop {} {\bibfield
  {journal} {\bibinfo  {journal} {Journal of Statistical Physics}\ }\textbf
  {\bibinfo {volume} {145}},\ \bibinfo {pages} {1256} (\bibinfo {year}
  {2011})}\BibitemShut {NoStop}%
\bibitem [{\citenamefont {Chirikov}\ and\ \citenamefont
  {Vecheslavov}(1997)}]{chirikov1997arnold}%
  \BibitemOpen
  \bibfield  {author} {\bibinfo {author} {\bibfnamefont {B.}~\bibnamefont
  {Chirikov}}\ and\ \bibinfo {author} {\bibfnamefont {V.}~\bibnamefont
  {Vecheslavov}},\ }\bibfield  {title} {\bibinfo {title} {Arnold diffusion in
  large systems},\ }\href@noop {} {\bibfield  {journal} {\bibinfo  {journal}
  {Journal of Experimental and Theoretical Physics}\ }\textbf {\bibinfo
  {volume} {85}},\ \bibinfo {pages} {616} (\bibinfo {year} {1997})}\BibitemShut
  {NoStop}%
\bibitem [{\citenamefont {Richter}\ \emph {et~al.}(2014)\citenamefont
  {Richter}, \citenamefont {Lange}, \citenamefont {B{\"a}cker},\ and\
  \citenamefont {Ketzmerick}}]{richter2014visualization}%
  \BibitemOpen
  \bibfield  {author} {\bibinfo {author} {\bibfnamefont {M.}~\bibnamefont
  {Richter}}, \bibinfo {author} {\bibfnamefont {S.}~\bibnamefont {Lange}},
  \bibinfo {author} {\bibfnamefont {A.}~\bibnamefont {B{\"a}cker}},\ and\
  \bibinfo {author} {\bibfnamefont {R.}~\bibnamefont {Ketzmerick}},\ }\bibfield
   {title} {\bibinfo {title} {Visualization and comparison of classical
  structures and quantum states of four-dimensional maps},\ }\href@noop {}
  {\bibfield  {journal} {\bibinfo  {journal} {Physical Review E}\ }\textbf
  {\bibinfo {volume} {89}},\ \bibinfo {pages} {022902} (\bibinfo {year}
  {2014})}\BibitemShut {NoStop}%
\bibitem [{\citenamefont {Citro}\ \emph {et~al.}(2015)\citenamefont {Citro},
  \citenamefont {Dalla~Torre}, \citenamefont {D’Alessio}, \citenamefont
  {Polkovnikov}, \citenamefont {Babadi}, \citenamefont {Oka},\ and\
  \citenamefont {Demler}}]{citro2015dynamical}%
  \BibitemOpen
  \bibfield  {author} {\bibinfo {author} {\bibfnamefont {R.}~\bibnamefont
  {Citro}}, \bibinfo {author} {\bibfnamefont {E.~G.}\ \bibnamefont
  {Dalla~Torre}}, \bibinfo {author} {\bibfnamefont {L.}~\bibnamefont
  {D’Alessio}}, \bibinfo {author} {\bibfnamefont {A.}~\bibnamefont
  {Polkovnikov}}, \bibinfo {author} {\bibfnamefont {M.}~\bibnamefont {Babadi}},
  \bibinfo {author} {\bibfnamefont {T.}~\bibnamefont {Oka}},\ and\ \bibinfo
  {author} {\bibfnamefont {E.}~\bibnamefont {Demler}},\ }\bibfield  {title}
  {\bibinfo {title} {Dynamical stability of a many-body kapitza pendulum},\
  }\href@noop {} {\bibfield  {journal} {\bibinfo  {journal} {Annals of
  Physics}\ }\textbf {\bibinfo {volume} {360}},\ \bibinfo {pages} {694}
  (\bibinfo {year} {2015})}\BibitemShut {NoStop}%
\bibitem [{\citenamefont {Rozenbaum}\ and\ \citenamefont
  {Galitski}(2017)}]{rozenbaum2017dynamical}%
  \BibitemOpen
  \bibfield  {author} {\bibinfo {author} {\bibfnamefont {E.~B.}\ \bibnamefont
  {Rozenbaum}}\ and\ \bibinfo {author} {\bibfnamefont {V.}~\bibnamefont
  {Galitski}},\ }\bibfield  {title} {\bibinfo {title} {Dynamical localization
  of coupled relativistic kicked rotors},\ }\href@noop {} {\bibfield  {journal}
  {\bibinfo  {journal} {Physical Review B}\ }\textbf {\bibinfo {volume} {95}},\
  \bibinfo {pages} {064303} (\bibinfo {year} {2017})}\BibitemShut {NoStop}%
\bibitem [{\citenamefont {Lellouch}\ \emph {et~al.}(2020)\citenamefont
  {Lellouch}, \citenamefont {Ran{\c{c}}on}, \citenamefont {De~Bi{\`e}vre},
  \citenamefont {Delande},\ and\ \citenamefont
  {Garreau}}]{lellouch2020dynamics}%
  \BibitemOpen
  \bibfield  {author} {\bibinfo {author} {\bibfnamefont {S.}~\bibnamefont
  {Lellouch}}, \bibinfo {author} {\bibfnamefont {A.}~\bibnamefont
  {Ran{\c{c}}on}}, \bibinfo {author} {\bibfnamefont {S.}~\bibnamefont
  {De~Bi{\`e}vre}}, \bibinfo {author} {\bibfnamefont {D.}~\bibnamefont
  {Delande}},\ and\ \bibinfo {author} {\bibfnamefont {J.~C.}\ \bibnamefont
  {Garreau}},\ }\bibfield  {title} {\bibinfo {title} {Dynamics of the
  mean-field-interacting quantum kicked rotor},\ }\href@noop {} {\bibfield
  {journal} {\bibinfo  {journal} {Physical Review A}\ }\textbf {\bibinfo
  {volume} {101}},\ \bibinfo {pages} {043624} (\bibinfo {year}
  {2020})}\BibitemShut {NoStop}%
\bibitem [{\citenamefont {Kundu}\ \emph {et~al.}(2021)\citenamefont {Kundu},
  \citenamefont {Rajak},\ and\ \citenamefont {Nag}}]{kundu2021dynamics}%
  \BibitemOpen
  \bibfield  {author} {\bibinfo {author} {\bibfnamefont {A.}~\bibnamefont
  {Kundu}}, \bibinfo {author} {\bibfnamefont {A.}~\bibnamefont {Rajak}},\ and\
  \bibinfo {author} {\bibfnamefont {T.}~\bibnamefont {Nag}},\ }\bibfield
  {title} {\bibinfo {title} {Dynamics of fluctuation correlation in a
  periodically driven classical system},\ }\href@noop {} {\bibfield  {journal}
  {\bibinfo  {journal} {Physical Review B}\ }\textbf {\bibinfo {volume}
  {104}},\ \bibinfo {pages} {075161} (\bibinfo {year} {2021})}\BibitemShut
  {NoStop}%
\bibitem [{\citenamefont {Haldar}\ \emph {et~al.}(2021)\citenamefont {Haldar},
  \citenamefont {Mu}, \citenamefont {Georgeot}, \citenamefont {Gong},
  \citenamefont {Miniatura},\ and\ \citenamefont
  {Lemari{\'e}}}]{haldar2021prethermalization}%
  \BibitemOpen
  \bibfield  {author} {\bibinfo {author} {\bibfnamefont {P.}~\bibnamefont
  {Haldar}}, \bibinfo {author} {\bibfnamefont {S.}~\bibnamefont {Mu}}, \bibinfo
  {author} {\bibfnamefont {B.}~\bibnamefont {Georgeot}}, \bibinfo {author}
  {\bibfnamefont {J.}~\bibnamefont {Gong}}, \bibinfo {author} {\bibfnamefont
  {C.}~\bibnamefont {Miniatura}},\ and\ \bibinfo {author} {\bibfnamefont
  {G.}~\bibnamefont {Lemari{\'e}}},\ }\bibfield  {title} {\bibinfo {title}
  {Prethermalization and wave condensation in a nonlinear disordered floquet
  system},\ }\href@noop {} {\bibfield  {journal} {\bibinfo  {journal} {arXiv
  preprint arXiv:2109.14347}\ } (\bibinfo {year} {2021})}\BibitemShut {NoStop}%
\bibitem [{\citenamefont {Russomanno}\ \emph {et~al.}(2021)\citenamefont
  {Russomanno}, \citenamefont {Fava},\ and\ \citenamefont
  {Fazio}}]{russomanno2021chaos}%
  \BibitemOpen
  \bibfield  {author} {\bibinfo {author} {\bibfnamefont {A.}~\bibnamefont
  {Russomanno}}, \bibinfo {author} {\bibfnamefont {M.}~\bibnamefont {Fava}},\
  and\ \bibinfo {author} {\bibfnamefont {R.}~\bibnamefont {Fazio}},\ }\bibfield
   {title} {\bibinfo {title} {Chaos and subdiffusion in infinite-range coupled
  quantum kicked rotors},\ }\href@noop {} {\bibfield  {journal} {\bibinfo
  {journal} {Physical Review B}\ }\textbf {\bibinfo {volume} {103}},\ \bibinfo
  {pages} {224301} (\bibinfo {year} {2021})}\BibitemShut {NoStop}%
\bibitem [{\citenamefont {Martinez}\ \emph {et~al.}(2022)\citenamefont
  {Martinez}, \citenamefont {Larr{\'e}}, \citenamefont {Delande},\ and\
  \citenamefont {Cherroret}}]{martinez2022low}%
  \BibitemOpen
  \bibfield  {author} {\bibinfo {author} {\bibfnamefont {M.}~\bibnamefont
  {Martinez}}, \bibinfo {author} {\bibfnamefont {P.-{\'E}.}\ \bibnamefont
  {Larr{\'e}}}, \bibinfo {author} {\bibfnamefont {D.}~\bibnamefont {Delande}},\
  and\ \bibinfo {author} {\bibfnamefont {N.}~\bibnamefont {Cherroret}},\
  }\bibfield  {title} {\bibinfo {title} {Low-energy prethermal phase and
  crossover to thermalization in nonlinear kicked rotors},\ }\href@noop {}
  {\bibfield  {journal} {\bibinfo  {journal} {Physical Review A}\ }\textbf
  {\bibinfo {volume} {106}},\ \bibinfo {pages} {043304} (\bibinfo {year}
  {2022})}\BibitemShut {NoStop}%
\bibitem [{\citenamefont {Cao}\ \emph {et~al.}(2022)\citenamefont {Cao},
  \citenamefont {Sajjad}, \citenamefont {Mas}, \citenamefont {Simmons},
  \citenamefont {Tanlimco}, \citenamefont {Nolasco-Martinez}, \citenamefont
  {Shimasaki}, \citenamefont {Kondakci}, \citenamefont {Galitski},\ and\
  \citenamefont {Weld}}]{cao2022interaction}%
  \BibitemOpen
  \bibfield  {author} {\bibinfo {author} {\bibfnamefont {A.}~\bibnamefont
  {Cao}}, \bibinfo {author} {\bibfnamefont {R.}~\bibnamefont {Sajjad}},
  \bibinfo {author} {\bibfnamefont {H.}~\bibnamefont {Mas}}, \bibinfo {author}
  {\bibfnamefont {E.~Q.}\ \bibnamefont {Simmons}}, \bibinfo {author}
  {\bibfnamefont {J.~L.}\ \bibnamefont {Tanlimco}}, \bibinfo {author}
  {\bibfnamefont {E.}~\bibnamefont {Nolasco-Martinez}}, \bibinfo {author}
  {\bibfnamefont {T.}~\bibnamefont {Shimasaki}}, \bibinfo {author}
  {\bibfnamefont {H.~E.}\ \bibnamefont {Kondakci}}, \bibinfo {author}
  {\bibfnamefont {V.}~\bibnamefont {Galitski}},\ and\ \bibinfo {author}
  {\bibfnamefont {D.~M.}\ \bibnamefont {Weld}},\ }\bibfield  {title} {\bibinfo
  {title} {Interaction-driven breakdown of dynamical localization in a kicked
  quantum gas},\ }\href@noop {} {\bibfield  {journal} {\bibinfo  {journal}
  {Nature Physics}\ }\textbf {\bibinfo {volume} {18}},\ \bibinfo {pages} {1302}
  (\bibinfo {year} {2022})}\BibitemShut {NoStop}%
\bibitem [{\citenamefont {Rajak}\ \emph {et~al.}(2019)\citenamefont {Rajak},
  \citenamefont {Dana},\ and\ \citenamefont
  {Dalla~Torre}}]{rajak2019characterizations}%
  \BibitemOpen
  \bibfield  {author} {\bibinfo {author} {\bibfnamefont {A.}~\bibnamefont
  {Rajak}}, \bibinfo {author} {\bibfnamefont {I.}~\bibnamefont {Dana}},\ and\
  \bibinfo {author} {\bibfnamefont {E.~G.}\ \bibnamefont {Dalla~Torre}},\
  }\bibfield  {title} {\bibinfo {title} {Characterizations of prethermal states
  in periodically driven many-body systems with unbounded chaotic diffusion},\
  }\href@noop {} {\bibfield  {journal} {\bibinfo  {journal} {Physical Review
  B}\ }\textbf {\bibinfo {volume} {100}},\ \bibinfo {pages} {100302} (\bibinfo
  {year} {2019})}\BibitemShut {NoStop}%
\bibitem [{\citenamefont {Wen}\ \emph {et~al.}(2021)\citenamefont {Wen},
  \citenamefont {Fan}, \citenamefont {Vishwanath},\ and\ \citenamefont
  {Gu}}]{wen2021periodically}%
  \BibitemOpen
  \bibfield  {author} {\bibinfo {author} {\bibfnamefont {X.}~\bibnamefont
  {Wen}}, \bibinfo {author} {\bibfnamefont {R.}~\bibnamefont {Fan}}, \bibinfo
  {author} {\bibfnamefont {A.}~\bibnamefont {Vishwanath}},\ and\ \bibinfo
  {author} {\bibfnamefont {Y.}~\bibnamefont {Gu}},\ }\bibfield  {title}
  {\bibinfo {title} {Periodically, quasiperiodically, and randomly driven
  conformal field theories},\ }\href@noop {} {\bibfield  {journal} {\bibinfo
  {journal} {Physical Review Research}\ }\textbf {\bibinfo {volume} {3}},\
  \bibinfo {pages} {023044} (\bibinfo {year} {2021})}\BibitemShut {NoStop}%
\bibitem [{\citenamefont {Timms}\ \emph {et~al.}(2021)\citenamefont {Timms},
  \citenamefont {Sieberer},\ and\ \citenamefont
  {Kolodrubetz}}]{timms2021quantized}%
  \BibitemOpen
  \bibfield  {author} {\bibinfo {author} {\bibfnamefont {C.~I.}\ \bibnamefont
  {Timms}}, \bibinfo {author} {\bibfnamefont {L.~M.}\ \bibnamefont
  {Sieberer}},\ and\ \bibinfo {author} {\bibfnamefont {M.~H.}\ \bibnamefont
  {Kolodrubetz}},\ }\bibfield  {title} {\bibinfo {title} {Quantized floquet
  topology with temporal noise},\ }\href@noop {} {\bibfield  {journal}
  {\bibinfo  {journal} {Physical Review Letters}\ }\textbf {\bibinfo {volume}
  {127}},\ \bibinfo {pages} {270601} (\bibinfo {year} {2021})}\BibitemShut
  {NoStop}%
\bibitem [{\citenamefont {Pilatowsky-Cameo}\ \emph {et~al.}(2023)\citenamefont
  {Pilatowsky-Cameo}, \citenamefont {Dag}, \citenamefont {Ho},\ and\
  \citenamefont {Choi}}]{pilatowskycameo2023complete}%
  \BibitemOpen
  \bibfield  {author} {\bibinfo {author} {\bibfnamefont {S.}~\bibnamefont
  {Pilatowsky-Cameo}}, \bibinfo {author} {\bibfnamefont {C.~B.}\ \bibnamefont
  {Dag}}, \bibinfo {author} {\bibfnamefont {W.~W.}\ \bibnamefont {Ho}},\ and\
  \bibinfo {author} {\bibfnamefont {S.}~\bibnamefont {Choi}},\ }\href@noop {}
  {\bibinfo {title} {Complete hilbert-space ergodicity in quantum dynamics of
  generalized fibonacci drives}} (\bibinfo {year} {2023}),\ \Eprint
  {https://arxiv.org/abs/2306.11792} {arXiv:2306.11792 [quant-ph]} \BibitemShut
  {NoStop}%
\bibitem [{\citenamefont {Verdeny}\ \emph {et~al.}(2016)\citenamefont
  {Verdeny}, \citenamefont {Puig},\ and\ \citenamefont
  {Mintert}}]{verdeny2016quasi}%
  \BibitemOpen
  \bibfield  {author} {\bibinfo {author} {\bibfnamefont {A.}~\bibnamefont
  {Verdeny}}, \bibinfo {author} {\bibfnamefont {J.}~\bibnamefont {Puig}},\ and\
  \bibinfo {author} {\bibfnamefont {F.}~\bibnamefont {Mintert}},\ }\bibfield
  {title} {\bibinfo {title} {Quasi-periodically driven quantum systems},\
  }\href@noop {} {\bibfield  {journal} {\bibinfo  {journal} {Zeitschrift
  f{\"u}r Naturforschung A}\ }\textbf {\bibinfo {volume} {71}},\ \bibinfo
  {pages} {897} (\bibinfo {year} {2016})}\BibitemShut {NoStop}%
\bibitem [{\citenamefont {Dumitrescu}\ \emph {et~al.}(2018)\citenamefont
  {Dumitrescu}, \citenamefont {Vasseur},\ and\ \citenamefont
  {Potter}}]{dumitrescu2018logarithmically}%
  \BibitemOpen
  \bibfield  {author} {\bibinfo {author} {\bibfnamefont {P.~T.}\ \bibnamefont
  {Dumitrescu}}, \bibinfo {author} {\bibfnamefont {R.}~\bibnamefont
  {Vasseur}},\ and\ \bibinfo {author} {\bibfnamefont {A.~C.}\ \bibnamefont
  {Potter}},\ }\bibfield  {title} {\bibinfo {title} {Logarithmically slow
  relaxation in quasiperiodically driven random spin chains},\ }\href@noop {}
  {\bibfield  {journal} {\bibinfo  {journal} {Physical review letters}\
  }\textbf {\bibinfo {volume} {120}},\ \bibinfo {pages} {070602} (\bibinfo
  {year} {2018})}\BibitemShut {NoStop}%
\bibitem [{\citenamefont {Mukherjee}\ \emph {et~al.}(2020)\citenamefont
  {Mukherjee}, \citenamefont {Sen}, \citenamefont {Sen},\ and\ \citenamefont
  {Sengupta}}]{mukherjee2020restoring}%
  \BibitemOpen
  \bibfield  {author} {\bibinfo {author} {\bibfnamefont {B.}~\bibnamefont
  {Mukherjee}}, \bibinfo {author} {\bibfnamefont {A.}~\bibnamefont {Sen}},
  \bibinfo {author} {\bibfnamefont {D.}~\bibnamefont {Sen}},\ and\ \bibinfo
  {author} {\bibfnamefont {K.}~\bibnamefont {Sengupta}},\ }\bibfield  {title}
  {\bibinfo {title} {Restoring coherence via aperiodic drives in a many-body
  quantum system},\ }\href@noop {} {\bibfield  {journal} {\bibinfo  {journal}
  {Physical Review B}\ }\textbf {\bibinfo {volume} {102}},\ \bibinfo {pages}
  {014301} (\bibinfo {year} {2020})}\BibitemShut {NoStop}%
\bibitem [{\citenamefont {Lapierre}\ \emph {et~al.}(2020)\citenamefont
  {Lapierre}, \citenamefont {Choo}, \citenamefont {Tiwari}, \citenamefont
  {Tauber}, \citenamefont {Neupert},\ and\ \citenamefont
  {Chitra}}]{lapierre2020fine}%
  \BibitemOpen
  \bibfield  {author} {\bibinfo {author} {\bibfnamefont {B.}~\bibnamefont
  {Lapierre}}, \bibinfo {author} {\bibfnamefont {K.}~\bibnamefont {Choo}},
  \bibinfo {author} {\bibfnamefont {A.}~\bibnamefont {Tiwari}}, \bibinfo
  {author} {\bibfnamefont {C.}~\bibnamefont {Tauber}}, \bibinfo {author}
  {\bibfnamefont {T.}~\bibnamefont {Neupert}},\ and\ \bibinfo {author}
  {\bibfnamefont {R.}~\bibnamefont {Chitra}},\ }\bibfield  {title} {\bibinfo
  {title} {Fine structure of heating in a quasiperiodically driven critical
  quantum system},\ }\href@noop {} {\bibfield  {journal} {\bibinfo  {journal}
  {Physical Review Research}\ }\textbf {\bibinfo {volume} {2}},\ \bibinfo
  {pages} {033461} (\bibinfo {year} {2020})}\BibitemShut {NoStop}%
\bibitem [{\citenamefont {Zhao}\ \emph {et~al.}(2021)\citenamefont {Zhao},
  \citenamefont {Mintert}, \citenamefont {Moessner},\ and\ \citenamefont
  {Knolle}}]{zhao2021random}%
  \BibitemOpen
  \bibfield  {author} {\bibinfo {author} {\bibfnamefont {H.}~\bibnamefont
  {Zhao}}, \bibinfo {author} {\bibfnamefont {F.}~\bibnamefont {Mintert}},
  \bibinfo {author} {\bibfnamefont {R.}~\bibnamefont {Moessner}},\ and\
  \bibinfo {author} {\bibfnamefont {J.}~\bibnamefont {Knolle}},\ }\bibfield
  {title} {\bibinfo {title} {Random multipolar driving: tunably slow heating
  through spectral engineering},\ }\href@noop {} {\bibfield  {journal}
  {\bibinfo  {journal} {Physical Review Letters}\ }\textbf {\bibinfo {volume}
  {126}},\ \bibinfo {pages} {040601} (\bibinfo {year} {2021})}\BibitemShut
  {NoStop}%
\bibitem [{\citenamefont {Cai}(2022)}]{cai20221}%
  \BibitemOpen
  \bibfield  {author} {\bibinfo {author} {\bibfnamefont {Z.}~\bibnamefont
  {Cai}},\ }\bibfield  {title} {\bibinfo {title} {1/3 power-law universality
  class out of stochastic driving in interacting systems},\ }\href@noop {}
  {\bibfield  {journal} {\bibinfo  {journal} {Physical Review Letters}\
  }\textbf {\bibinfo {volume} {128}},\ \bibinfo {pages} {050601} (\bibinfo
  {year} {2022})}\BibitemShut {NoStop}%
\bibitem [{\citenamefont {Ying}\ \emph {et~al.}(2022)\citenamefont {Ying},
  \citenamefont {Guo}, \citenamefont {Li}, \citenamefont {Gong}, \citenamefont
  {Deng}, \citenamefont {Chen}, \citenamefont {Zha}, \citenamefont {Ye},
  \citenamefont {Wang}, \citenamefont {Zhu} \emph {et~al.}}]{ying2022floquet}%
  \BibitemOpen
  \bibfield  {author} {\bibinfo {author} {\bibfnamefont {C.}~\bibnamefont
  {Ying}}, \bibinfo {author} {\bibfnamefont {Q.}~\bibnamefont {Guo}}, \bibinfo
  {author} {\bibfnamefont {S.}~\bibnamefont {Li}}, \bibinfo {author}
  {\bibfnamefont {M.}~\bibnamefont {Gong}}, \bibinfo {author} {\bibfnamefont
  {X.-H.}\ \bibnamefont {Deng}}, \bibinfo {author} {\bibfnamefont
  {F.}~\bibnamefont {Chen}}, \bibinfo {author} {\bibfnamefont {C.}~\bibnamefont
  {Zha}}, \bibinfo {author} {\bibfnamefont {Y.}~\bibnamefont {Ye}}, \bibinfo
  {author} {\bibfnamefont {C.}~\bibnamefont {Wang}}, \bibinfo {author}
  {\bibfnamefont {Q.}~\bibnamefont {Zhu}}, \emph {et~al.},\ }\bibfield  {title}
  {\bibinfo {title} {Floquet prethermal phase protected by u (1) symmetry on a
  superconducting quantum processor},\ }\href@noop {} {\bibfield  {journal}
  {\bibinfo  {journal} {Physical Review A}\ }\textbf {\bibinfo {volume}
  {105}},\ \bibinfo {pages} {012418} (\bibinfo {year} {2022})}\BibitemShut
  {NoStop}%
\bibitem [{\citenamefont {He}\ \emph {et~al.}(2022)\citenamefont {He},
  \citenamefont {Ye}, \citenamefont {Gong}, \citenamefont {Liu}, \citenamefont
  {Murch}, \citenamefont {Yao},\ and\ \citenamefont {Zu}}]{he2022quasi}%
  \BibitemOpen
  \bibfield  {author} {\bibinfo {author} {\bibfnamefont {G.}~\bibnamefont
  {He}}, \bibinfo {author} {\bibfnamefont {B.}~\bibnamefont {Ye}}, \bibinfo
  {author} {\bibfnamefont {R.}~\bibnamefont {Gong}}, \bibinfo {author}
  {\bibfnamefont {Z.}~\bibnamefont {Liu}}, \bibinfo {author} {\bibfnamefont
  {K.~W.}\ \bibnamefont {Murch}}, \bibinfo {author} {\bibfnamefont {N.~Y.}\
  \bibnamefont {Yao}},\ and\ \bibinfo {author} {\bibfnamefont {C.}~\bibnamefont
  {Zu}},\ }\bibfield  {title} {\bibinfo {title} {Quasi-floquet
  prethermalization in a disordered dipolar spin ensemble in diamond},\
  }\href@noop {} {\bibfield  {journal} {\bibinfo  {journal} {arXiv preprint
  arXiv:2212.11284}\ } (\bibinfo {year} {2022})}\BibitemShut {NoStop}%
\bibitem [{\citenamefont {Long}\ \emph {et~al.}(2022)\citenamefont {Long},
  \citenamefont {Crowley},\ and\ \citenamefont {Chandran}}]{long2022many}%
  \BibitemOpen
  \bibfield  {author} {\bibinfo {author} {\bibfnamefont {D.~M.}\ \bibnamefont
  {Long}}, \bibinfo {author} {\bibfnamefont {P.~J.}\ \bibnamefont {Crowley}},\
  and\ \bibinfo {author} {\bibfnamefont {A.}~\bibnamefont {Chandran}},\
  }\bibfield  {title} {\bibinfo {title} {Many-body localization with
  quasiperiodic driving},\ }\href@noop {} {\bibfield  {journal} {\bibinfo
  {journal} {Physical Review B}\ }\textbf {\bibinfo {volume} {105}},\ \bibinfo
  {pages} {144204} (\bibinfo {year} {2022})}\BibitemShut {NoStop}%
\bibitem [{\citenamefont {Martin}\ \emph {et~al.}(2022)\citenamefont {Martin},
  \citenamefont {Martin},\ and\ \citenamefont {Agarwal}}]{martin2022effect}%
  \BibitemOpen
  \bibfield  {author} {\bibinfo {author} {\bibfnamefont {T.}~\bibnamefont
  {Martin}}, \bibinfo {author} {\bibfnamefont {I.}~\bibnamefont {Martin}},\
  and\ \bibinfo {author} {\bibfnamefont {K.}~\bibnamefont {Agarwal}},\
  }\bibfield  {title} {\bibinfo {title} {Effect of quasiperiodic and random
  noise on many-body dynamical decoupling protocols},\ }\href@noop {}
  {\bibfield  {journal} {\bibinfo  {journal} {Physical Review B}\ }\textbf
  {\bibinfo {volume} {106}},\ \bibinfo {pages} {134306} (\bibinfo {year}
  {2022})}\BibitemShut {NoStop}%
\bibitem [{\citenamefont {Zhao}\ \emph {et~al.}(2022)\citenamefont {Zhao},
  \citenamefont {Knolle},\ and\ \citenamefont {Moessner}}]{zhao2022temporal}%
  \BibitemOpen
  \bibfield  {author} {\bibinfo {author} {\bibfnamefont {H.}~\bibnamefont
  {Zhao}}, \bibinfo {author} {\bibfnamefont {J.}~\bibnamefont {Knolle}},\ and\
  \bibinfo {author} {\bibfnamefont {R.}~\bibnamefont {Moessner}},\ }\bibfield
  {title} {\bibinfo {title} {Temporal disorder in spatiotemporal order},\
  }\href@noop {} {\bibfield  {journal} {\bibinfo  {journal} {arXiv preprint
  arXiv:2212.03135}\ } (\bibinfo {year} {2022})}\BibitemShut {NoStop}%
\bibitem [{\citenamefont {Tiwari}\ \emph {et~al.}(2023)\citenamefont {Tiwari},
  \citenamefont {Bhakuni},\ and\ \citenamefont {Sharma}}]{tiwari2023dynamical}%
  \BibitemOpen
  \bibfield  {author} {\bibinfo {author} {\bibfnamefont {V.}~\bibnamefont
  {Tiwari}}, \bibinfo {author} {\bibfnamefont {D.~S.}\ \bibnamefont
  {Bhakuni}},\ and\ \bibinfo {author} {\bibfnamefont {A.}~\bibnamefont
  {Sharma}},\ }\bibfield  {title} {\bibinfo {title} {Dynamical localization and
  slow dynamics in quasiperiodically-driven quantum systems},\ }\href@noop {}
  {\bibfield  {journal} {\bibinfo  {journal} {arXiv preprint arXiv:2302.12271}\
  } (\bibinfo {year} {2023})}\BibitemShut {NoStop}%
\bibitem [{\citenamefont {Else}\ \emph {et~al.}(2020)\citenamefont {Else},
  \citenamefont {Ho},\ and\ \citenamefont {Dumitrescu}}]{else2020long}%
  \BibitemOpen
  \bibfield  {author} {\bibinfo {author} {\bibfnamefont {D.~V.}\ \bibnamefont
  {Else}}, \bibinfo {author} {\bibfnamefont {W.~W.}\ \bibnamefont {Ho}},\ and\
  \bibinfo {author} {\bibfnamefont {P.~T.}\ \bibnamefont {Dumitrescu}},\
  }\bibfield  {title} {\bibinfo {title} {Long-lived interacting phases of
  matter protected by multiple time-translation symmetries in quasiperiodically
  driven systems},\ }\href@noop {} {\bibfield  {journal} {\bibinfo  {journal}
  {Physical Review X}\ }\textbf {\bibinfo {volume} {10}},\ \bibinfo {pages}
  {021032} (\bibinfo {year} {2020})}\BibitemShut {NoStop}%
\bibitem [{\citenamefont {Mori}\ \emph {et~al.}(2021)\citenamefont {Mori},
  \citenamefont {Zhao}, \citenamefont {Mintert}, \citenamefont {Knolle},\ and\
  \citenamefont {Moessner}}]{mori2021rigorous}%
  \BibitemOpen
  \bibfield  {author} {\bibinfo {author} {\bibfnamefont {T.}~\bibnamefont
  {Mori}}, \bibinfo {author} {\bibfnamefont {H.}~\bibnamefont {Zhao}}, \bibinfo
  {author} {\bibfnamefont {F.}~\bibnamefont {Mintert}}, \bibinfo {author}
  {\bibfnamefont {J.}~\bibnamefont {Knolle}},\ and\ \bibinfo {author}
  {\bibfnamefont {R.}~\bibnamefont {Moessner}},\ }\bibfield  {title} {\bibinfo
  {title} {Rigorous bounds on the heating rate in thue-morse quasiperiodically
  and randomly driven quantum many-body systems},\ }\href@noop {} {\bibfield
  {journal} {\bibinfo  {journal} {Physical Review Letters}\ }\textbf {\bibinfo
  {volume} {127}},\ \bibinfo {pages} {050602} (\bibinfo {year}
  {2021})}\BibitemShut {NoStop}%
\bibitem [{\citenamefont {Casati}\ \emph {et~al.}(1989)\citenamefont {Casati},
  \citenamefont {Guarneri},\ and\ \citenamefont
  {Shepelyansky}}]{casati1989anderson}%
  \BibitemOpen
  \bibfield  {author} {\bibinfo {author} {\bibfnamefont {G.}~\bibnamefont
  {Casati}}, \bibinfo {author} {\bibfnamefont {I.}~\bibnamefont {Guarneri}},\
  and\ \bibinfo {author} {\bibfnamefont {D.}~\bibnamefont {Shepelyansky}},\
  }\bibfield  {title} {\bibinfo {title} {Anderson transition in a
  one-dimensional system with three incommensurate frequencies},\ }\href@noop
  {} {\bibfield  {journal} {\bibinfo  {journal} {Physical review letters}\
  }\textbf {\bibinfo {volume} {62}},\ \bibinfo {pages} {345} (\bibinfo {year}
  {1989})}\BibitemShut {NoStop}%
\bibitem [{\citenamefont {Lemari{\'e}}\ \emph {et~al.}(2009)\citenamefont
  {Lemari{\'e}}, \citenamefont {Gr{\'e}maud},\ and\ \citenamefont
  {Delande}}]{lemarie2009universality}%
  \BibitemOpen
  \bibfield  {author} {\bibinfo {author} {\bibfnamefont {G.}~\bibnamefont
  {Lemari{\'e}}}, \bibinfo {author} {\bibfnamefont {B.}~\bibnamefont
  {Gr{\'e}maud}},\ and\ \bibinfo {author} {\bibfnamefont {D.}~\bibnamefont
  {Delande}},\ }\bibfield  {title} {\bibinfo {title} {Universality of the
  anderson transition with the quasiperiodic kicked rotor},\ }\href@noop {}
  {\bibfield  {journal} {\bibinfo  {journal} {Europhysics Letters}\ }\textbf
  {\bibinfo {volume} {87}},\ \bibinfo {pages} {37007} (\bibinfo {year}
  {2009})}\BibitemShut {NoStop}%
\bibitem [{\citenamefont {Goldfriend}\ and\ \citenamefont
  {Kurchan}(2020)}]{goldfriend2020quasi}%
  \BibitemOpen
  \bibfield  {author} {\bibinfo {author} {\bibfnamefont {T.}~\bibnamefont
  {Goldfriend}}\ and\ \bibinfo {author} {\bibfnamefont {J.}~\bibnamefont
  {Kurchan}},\ }\bibfield  {title} {\bibinfo {title} {Quasi-integrable systems
  are slow to thermalize but may be good scramblers},\ }\href@noop {}
  {\bibfield  {journal} {\bibinfo  {journal} {Physical Review E}\ }\textbf
  {\bibinfo {volume} {102}},\ \bibinfo {pages} {022201} (\bibinfo {year}
  {2020})}\BibitemShut {NoStop}%
\bibitem [{\citenamefont {Santhanam}\ \emph {et~al.}(2022)\citenamefont
  {Santhanam}, \citenamefont {Paul},\ and\ \citenamefont
  {Kannan}}]{santhanam2022quantum}%
  \BibitemOpen
  \bibfield  {author} {\bibinfo {author} {\bibfnamefont {M.}~\bibnamefont
  {Santhanam}}, \bibinfo {author} {\bibfnamefont {S.}~\bibnamefont {Paul}},\
  and\ \bibinfo {author} {\bibfnamefont {J.~B.}\ \bibnamefont {Kannan}},\
  }\bibfield  {title} {\bibinfo {title} {Quantum kicked rotor and its variants:
  Chaos, localization and beyond},\ }\href@noop {} {\bibfield  {journal}
  {\bibinfo  {journal} {Physics Reports}\ }\textbf {\bibinfo {volume} {956}},\
  \bibinfo {pages} {1} (\bibinfo {year} {2022})}\BibitemShut {NoStop}%
\bibitem [{\citenamefont {Vuatelet}\ and\ \citenamefont
  {Ran{\c{c}}on}(2023)}]{vuatelet2023dynamical}%
  \BibitemOpen
  \bibfield  {author} {\bibinfo {author} {\bibfnamefont {V.}~\bibnamefont
  {Vuatelet}}\ and\ \bibinfo {author} {\bibfnamefont {A.}~\bibnamefont
  {Ran{\c{c}}on}},\ }\bibfield  {title} {\bibinfo {title} {Dynamical many-body
  delocalization transition of a tonks gas in a quasi-periodic driving
  potential},\ }\href@noop {} {\bibfield  {journal} {\bibinfo  {journal}
  {Quantum}\ }\textbf {\bibinfo {volume} {7}},\ \bibinfo {pages} {917}
  (\bibinfo {year} {2023})}\BibitemShut {NoStop}%
\bibitem [{\citenamefont {Nandy}\ \emph {et~al.}(2017)\citenamefont {Nandy},
  \citenamefont {Sen},\ and\ \citenamefont {Sen}}]{nandy2017aperiodically}%
  \BibitemOpen
  \bibfield  {author} {\bibinfo {author} {\bibfnamefont {S.}~\bibnamefont
  {Nandy}}, \bibinfo {author} {\bibfnamefont {A.}~\bibnamefont {Sen}},\ and\
  \bibinfo {author} {\bibfnamefont {D.}~\bibnamefont {Sen}},\ }\bibfield
  {title} {\bibinfo {title} {Aperiodically driven integrable systems and their
  emergent steady states},\ }\href@noop {} {\bibfield  {journal} {\bibinfo
  {journal} {Physical Review X}\ }\textbf {\bibinfo {volume} {7}},\ \bibinfo
  {pages} {031034} (\bibinfo {year} {2017})}\BibitemShut {NoStop}%
\bibitem [{Note1()}]{Note1}%
  \BibitemOpen
  \bibinfo {note} {$N=500$ is chosen to sufficiently mimic the heating behavior
  in thermodynamically large systems; further details can be found in Sec.~\ref
  {sec.finite_size}.}\BibitemShut {Stop}%
\bibitem [{\citenamefont {Kruscha}\ \emph {et~al.}(2012)\citenamefont
  {Kruscha}, \citenamefont {Ketzmerick},\ and\ \citenamefont
  {Kantz}}]{kruscha2012biased}%
  \BibitemOpen
  \bibfield  {author} {\bibinfo {author} {\bibfnamefont {A.}~\bibnamefont
  {Kruscha}}, \bibinfo {author} {\bibfnamefont {R.}~\bibnamefont
  {Ketzmerick}},\ and\ \bibinfo {author} {\bibfnamefont {H.}~\bibnamefont
  {Kantz}},\ }\bibfield  {title} {\bibinfo {title} {Biased diffusion inside
  regular islands under random symplectic perturbations},\ }\href@noop {}
  {\bibfield  {journal} {\bibinfo  {journal} {Physical Review E}\ }\textbf
  {\bibinfo {volume} {85}},\ \bibinfo {pages} {066210} (\bibinfo {year}
  {2012})}\BibitemShut {NoStop}%
\bibitem [{Note2()}]{Note2}%
  \BibitemOpen
  \bibinfo {note} {The area is given by the expression \begin {align*} A(Q,
  P)=\protect \frac {\pi \left [M_{12} P^2-M_{21} Q^2+\left
  (M_{11}-M_{22}\right ) Q P\right ]}{\protect \sqrt {1-\left (\protect \frac
  {M_{11}+M_{22}}{2}\right )^2}}, \end {align*} as a function of $Q$ and $P$
  and $M_{ij}$ denotes the matrix elements of $\protect \bar {M}'$~\cite
  {lichtenberg2013regular}.}\BibitemShut {Stop}%
\bibitem [{\citenamefont {Cataliotti}\ \emph {et~al.}(2001)\citenamefont
  {Cataliotti}, \citenamefont {Burger}, \citenamefont {Fort}, \citenamefont
  {Maddaloni}, \citenamefont {Minardi}, \citenamefont {Trombettoni},
  \citenamefont {Smerzi},\ and\ \citenamefont
  {Inguscio}}]{cataliotti2001josephson}%
  \BibitemOpen
  \bibfield  {author} {\bibinfo {author} {\bibfnamefont {F.}~\bibnamefont
  {Cataliotti}}, \bibinfo {author} {\bibfnamefont {S.}~\bibnamefont {Burger}},
  \bibinfo {author} {\bibfnamefont {C.}~\bibnamefont {Fort}}, \bibinfo {author}
  {\bibfnamefont {P.}~\bibnamefont {Maddaloni}}, \bibinfo {author}
  {\bibfnamefont {F.}~\bibnamefont {Minardi}}, \bibinfo {author} {\bibfnamefont
  {A.}~\bibnamefont {Trombettoni}}, \bibinfo {author} {\bibfnamefont
  {A.}~\bibnamefont {Smerzi}},\ and\ \bibinfo {author} {\bibfnamefont
  {M.}~\bibnamefont {Inguscio}},\ }\bibfield  {title} {\bibinfo {title}
  {Josephson junction arrays with bose-einstein condensates},\ }\href@noop {}
  {\bibfield  {journal} {\bibinfo  {journal} {Science}\ }\textbf {\bibinfo
  {volume} {293}},\ \bibinfo {pages} {843} (\bibinfo {year}
  {2001})}\BibitemShut {NoStop}%
\bibitem [{\citenamefont {Bloch}\ \emph {et~al.}(2008)\citenamefont {Bloch},
  \citenamefont {Dalibard},\ and\ \citenamefont {Zwerger}}]{bloch2008many}%
  \BibitemOpen
  \bibfield  {author} {\bibinfo {author} {\bibfnamefont {I.}~\bibnamefont
  {Bloch}}, \bibinfo {author} {\bibfnamefont {J.}~\bibnamefont {Dalibard}},\
  and\ \bibinfo {author} {\bibfnamefont {W.}~\bibnamefont {Zwerger}},\
  }\bibfield  {title} {\bibinfo {title} {Many-body physics with ultracold
  gases},\ }\href@noop {} {\bibfield  {journal} {\bibinfo  {journal} {Reviews
  of modern physics}\ }\textbf {\bibinfo {volume} {80}},\ \bibinfo {pages}
  {885} (\bibinfo {year} {2008})}\BibitemShut {NoStop}%
\bibitem [{Note3()}]{Note3}%
  \BibitemOpen
  \bibinfo {note} {Note that simulating the time evolution with a static
  interaction requires discretizing the continuous equations of motion, which
  significantly increases the numerical cost for long simulations of the
  dynamics. Instead, we modify the kick amplitude of the interaction $V$ as
  $K_l=\pm K+B$, allowing us to generate this interaction approximately at
  stroboscopic times.}\BibitemShut {Stop}%
\bibitem [{\citenamefont {Rajak}\ \emph {et~al.}(2018)\citenamefont {Rajak},
  \citenamefont {Citro},\ and\ \citenamefont
  {Dalla~Torre}}]{rajak2018stability}%
  \BibitemOpen
  \bibfield  {author} {\bibinfo {author} {\bibfnamefont {A.}~\bibnamefont
  {Rajak}}, \bibinfo {author} {\bibfnamefont {R.}~\bibnamefont {Citro}},\ and\
  \bibinfo {author} {\bibfnamefont {E.~G.}\ \bibnamefont {Dalla~Torre}},\
  }\bibfield  {title} {\bibinfo {title} {Stability and pre-thermalization in
  chains of classical kicked rotors},\ }\href@noop {} {\bibfield  {journal}
  {\bibinfo  {journal} {Journal of Physics A: Mathematical and Theoretical}\
  }\textbf {\bibinfo {volume} {51}},\ \bibinfo {pages} {465001} (\bibinfo
  {year} {2018})}\BibitemShut {NoStop}%
\bibitem [{\citenamefont {Lichtenberg}\ and\ \citenamefont
  {Lieberman}(2013)}]{lichtenberg2013regular}%
  \BibitemOpen
  \bibfield  {author} {\bibinfo {author} {\bibfnamefont {A.~J.}\ \bibnamefont
  {Lichtenberg}}\ and\ \bibinfo {author} {\bibfnamefont {M.~A.}\ \bibnamefont
  {Lieberman}},\ }\href@noop {} {\emph {\bibinfo {title} {Regular and chaotic
  dynamics}}},\ Vol.~\bibinfo {volume} {38}\ (\bibinfo  {publisher} {Springer
  Science \& Business Media},\ \bibinfo {year} {2013})\BibitemShut {NoStop}%
\end{thebibliography}%
 \let\addcontentsline\oldaddcontentsline 
 
	\cleardoublepage
	\onecolumngrid

\begin{center}
		\textbf{\large{\textit{Supplementary Material} \\ \smallskip
	Prethermalization in aperiodically kicked many-body dynamics}}\\
		\hfill \break
		\smallskip
\end{center}
	
	\renewcommand{\thefigure}{S\arabic{figure}}
	\setcounter{figure}{0}
	\renewcommand{\theequation}{S.\arabic{equation}}
	\setcounter{equation}{0}
	\renewcommand{\thesection}{SM\;\arabic{section}}
	\setcounter{section}{0}
	\tableofcontents

\section{Many-body effective Hamiltonian for kicked systems}
\label{sec:perturbative_expansion}
For the quantum kicked system, we have two different unitary time evolution operators
\begin{equation}
    U_0^{+}=e^{-i\tau H}e^{-iKV},\\
    U_0^{-}=e^{-i\tau H}e^{iKV},\\
\end{equation}
One can obtain a time-averaged Hamiltonian $H_{\mathrm{ave}}^{\pm}=H \pm \frac{K}{\tau}V$ as the effective Hamiltonian to approximate the early time dynamics. However, this is not a suitable expansion if $\tau$ is not sufficiently small and one wants to study the perturbative expansion with respect to the kick strength $K$. For instance, the term $[H,V]_s$ all have an amplitude scaling of $\mathcal{O}(K)$ but they are not captured in the averaged Hamiltonian. Instead, it is necessary to perform the replica resummation, whose general expression can be cumbersome to obtain, but there is a systematic approach to achieve it~\cite{vajna2018replica}, see also examples in \cite{fleckenstein2021thermalization}. 
For the same reason, the previous heating analysis on RMD systems in the high-frequency regime and the expansion of order $\mathcal{O}(\tau)$ will not be applicable here.

To explore the heating effect in RMD kicked systems, we consider the expansion in $\mathcal{O}(K)$ as $H^{\pm}_{n}=\sum_{m=0}^{\infty}K^m\Omega_{n,m}^{\pm}$ such that $U_n^{\pm}=\exp(-i2^n\tau H_n^{\pm})$. The symmetry $H^{+}_n\to H^-_n$ (under $K\to-K$) implies that 
\begin{equation}
\label{eq.symmetry_SM}
     \Omega_{n,m}^{+}=(-1)^m\Omega_{n,m}^-\coloneqq \Omega_{n,m}
\end{equation} 
for all $n$.
For $n=0$, a systematic method has been established for constructing the expansion
$U_0^{\pm}=\exp[-i\tau (H_0\pm K\Omega_{0,1}+\mathcal{O}(K^2))]$
and the $\mathcal{O}(K)$ term is presented in Eq.~\ref{eq.expansion_n0} ~\cite{vajna2018replica} . For larger values of $n$ we still begin by considering the leading-order correction with $m=1$, and we assume a general structure for $n\ge 1$ as
\begin{equation}
    \label{eq.ansatz}
\Omega_{n,1}^{\pm}=\sum_{s=0}^{\infty}f_{n,s}^{\pm}\tau^s [H,\Omega_{0,1}]_s, \ \  [H,\Omega_{0,1}]_s \coloneqq [\underbrace{H,\dots,[H}_\text{s},\Omega_{0,1}]\dots].
\end{equation} From Eq.~\ref{eq.symmetry_SM} we know for any $n$ and $s$, we have $f_{n,s}^{+}=-f_{n,s}^{-}$ for $m=1$. 
Note, we do not require the specific expression for each coefficient $f_{n,s}^{\pm}$. Instead, it suffices to demonstrate that some of them become exactly zero, thereby prohibiting certain heating channels. A similar expansion can be derived for higher-order multipolar operators
\begin{equation}
  \begin{aligned}  U_{n+1}^{\mp}=U_n^{\pm}U_n^{\mp}=&\exp\Big\{-i2^{n+1}\tau\Big[H+K(f_{n,s}^{\pm}+f_{n,s}^{\mp})\sum_{s=0}^{\infty}\tau^s [H,\Omega_{0,1}]_s/2
  \\
 &- (-i2^{n-1})K\sum_{s=0}^{\infty}\tau^{s+1}f_{n,s}^{\pm}[H,\Omega_{0,1}]_{s+1}
+K\sum_{l=2}^{\infty}\sum_{s=0}^{\infty}\tau^{s+l}g_{n,s,l}^{\pm}[H,\Omega_{0,1}]_{s+l}+\mathcal{O}(K^2)\Big]\Big\}     \end{aligned},
\end{equation}
where $g_{n,s,l}$ are some coefficients, and importantly, $f_{n,s}^{\pm}+f_{n,s}^{\mp}=0$ in the first line cancels. 
By comparing it with the assumption Eq.~\ref{eq.ansatz} but for $n=1$, we have 
\begin{equation}
  \begin{aligned}  U_{n+1}^{\mp}
  =&\exp\left\{-i2^{n+1}\tau\left[H+K\sum_{s=0}^{\infty}\tau^{s}f_{n+1,s}^{\mp}[H,\Omega_{0,1}]_s\right]+\mathcal{O}(K^2)\right\} ,
    \end{aligned}
\end{equation}
and by matching the coefficients of $\tau^s$, we can establish the following relation for the coefficients
\begin{equation}
    \begin{aligned}
    \label{eq.coefficient}
        &f_{n,s}^{+}=0, \text{ for}\  s\le n-1,\\
        &f_{n,s}^{+}=(-i)2^{n-2}f_{n-1,s-1}^+, \text{ for}\ s=n,
    \end{aligned}
\end{equation}
and obtaining $f_{n,s}^{+}$ for $s\ge n+1$ can be cumbersome. Therefore, the first line in Eq.~\ref{eq.coefficient} implies Eq.~\ref{eq.ansatz_main} in the main text. It suggests that, to the leading order of $\mathcal{O}(K)$, heating can only occur via the process $[H,V]_s$ with $s\ge n$, while all other heating channels are strictly forbidden. 

Higher-order terms with an even order of $K$ do not introduce random perturbations. However, they still contribute to heating in the form of Arnold diffusion, similar to periodically driven systems, but their contribution is exponentially small in the kick strength~\cite{rajak2019characterizations}. As a result, the next significant random heating channels emerge at order $\mathcal{O}(K^3)$. It can also be shown that the self-similarity of the RMD sequence leads to the exact suppression of these heating channels. To see this more easily, we consider a special case where higher-order nested commutators have negligible contributions 
\begin{equation}
   [H,V]_s=0,  
\end{equation}
for $\forall s\ge n_c$ with a certain integer $n_c$. Consequently, $\Omega_{n,1}=0$ and the stroboscopic time evolution of the system is effectively governed by the Hamiltonian $H^{\pm}_n=H+K^2\Omega_{n,2}\pm K^3\Omega_{n,3}+\mathcal{O}(K^4)$, where
\begin{equation}
    \label{eq.ansatz_2}
\Omega_{n,3}=\sum_{s=0}^{\infty}h_{n,s}\tau^s[H,\Omega_{k,3}]_s,
\end{equation}
for $n> k$ and a certain integer $k$. By using $U_{n+1}^{\mp}=U_n^{\pm}U_n^{\mp}$, one can again observe the vanishing coefficients 
\begin{equation}
    h_{n,s}=0, \text{ for}\ s\leq n-k-1, \quad n\geq k+1.
\end{equation}

\section{Linearization of the many-body Hamiltonian}
\subsection{Time evolution matrix}
\label{sec.linearization}
Following \cite{rajak2018stability}, we can express the Hamiltonian of our model as a collection of decoupled kicked harmonic oscillators in a quadratic approximation: $\cos (q_j - q_{j+1}) \approx 1 - (q_j - q_{j+1})^2/2$, provided that the two neighbouring rotor angles are sufficiently close $(q_j - q_{j+1}) \text{ mod } 2\pi \approx 0$. Thus, we have 
\begin{equation}
\begin{split}
H(t) &= \sum_{j=1}^N \left[ \frac{p_j^2}{2} - (B \pm K) \cos (q_j - q_{j+1}) \sum_{l = -\infty}^{+\infty} \delta (t - l\tau) \right] \\
&= \frac{1}{2}\sum_w \left[ |P_w|^2 + F^{\pm}(w) |Q_w|^2 \sum_{l = -\infty}^{+\infty}\delta (t - l\tau) \right], 
\end{split}
\end{equation}
where $w := 2\pi I/N$, $F^{\pm}(w) := 4(B \pm K)\sin^2(w/2)$ (the choice  of $F^{\pm}(w)$ depends on the RMD sequence), $P_w = \frac{1}{\sqrt{N}}\sum_{j = 1}^N p_j e^{-iwj}$ and $Q_w = \frac{1}{\sqrt{N}}\sum_{j = 1}^N q_j e^{-iwj}$ are the Fourier transforms of $p_j$ and $q_j$, respectively. Note that Eq.~\ref{eq.decouplingHamiltonian} in the main text is a simplified version of the Hamiltonian above with $B=0$. Here we use the general expression with non-zero $B$ such that the linear stability of the dynamics in Fig.~\ref{fig-B0pt01} can also be discussed.

Since $p_j$ and $q_j$ are real, we have $P^*_w = P_{-w}$ and $Q^*_w = Q_{-w}$ (here star denotes complex conjugate). For each $w$, the classical equations of motion are given by 
\begin{equation}
\frac{d}{dt}
\begin{pmatrix} Q_w \\ Q_{-w} \\ P_w \\ P_{-w} \end{pmatrix} 
= 
\begin{pmatrix} P_{-w} \\ P_w \\ -F^{\pm}(w)\Omega(t) Q_{-w} \\ -F^{\pm}(w)\Omega(t) Q_w \end{pmatrix} = \begin{pmatrix} 
0 & 0 & 0 & 1 \\
0 & 0 & 1 & 0 \\
0 & -F^{\pm}(w)\Omega(t) & 0 & 0 \\
-F^{\pm}(w)\Omega(t) & 0 & 0 & 0
\end{pmatrix} 
\begin{pmatrix} 
Q_w \\ Q_{-w} \\ P_w \\ P_{-w} 
\end{pmatrix} =: M^{\pm}(t) \begin{pmatrix} 
Q_w \\ Q_{-w} \\ P_w \\ P_{-w} 
\end{pmatrix},  
\end{equation}
where $\Omega(t) = \sum_{k = -\infty}^{+\infty} \delta (t - k\tau)$. 
\\
Consider the evolution of the system over one time period, from $t = -\epsilon$ to $t = \tau - \epsilon$ with $\epsilon \ll \tau$. The solution to the above equation is 
\begin{equation}
\begin{pmatrix} 
Q_w(\tau-\epsilon) \\
Q_{-w}(\tau-\epsilon) \\
P_w(\tau-\epsilon) \\
P_{-w}(\tau-\epsilon)
\end{pmatrix} \sim 
\exp \left[ \int_{-\epsilon}^{\tau-\epsilon} M^{\pm}(t)dt \right] 
\begin{pmatrix} 
Q_w(-\epsilon)  \\
Q_{-w}(-\epsilon)  \\
P_w(-\epsilon) \\
P_{-w}(-\epsilon)
\end{pmatrix}. 
\end{equation}
During the first part of the period (when $t \in (-\epsilon, \epsilon)$), the rotor is kicked and the time evolution is determined by the matrix 
\begin{equation}
M^{\pm}_{\text{kick}, w} = \lim_{\epsilon \to 0} \exp \left[ \int_{-\epsilon}^{+\epsilon} M^{\pm}(t)dt \right] = \exp \begin{pmatrix} 
0 & 0 & 0 & 0 \\
0 & 0 & 0 & 0 \\
0 & -F^{\pm}(w) & 0 & 0 \\
-F^{\pm}(w) & 0 & 0 & 0
\end{pmatrix} 
= \begin{pmatrix} 
1 & 0 & 0 & 0 \\
0 & 1 & 0 & 0 \\
0 & -F^{\pm}(w) & 1 & 0 \\
-F^{\pm}(w) & 0 & 0 & 1
\end{pmatrix}.
\end{equation}
In the second part of the time period (when $t \in (\epsilon, \tau-\epsilon)$), the rotor experiences a free motion, described by the matrix (with $\epsilon \to 0$) 
\begin{equation}
M_{\text{free}} = 
\begin{pmatrix} 
1 & 0 & 0 & \tau \\
0 & 1 & \tau & 0 \\
0 & 0 & 1 & 0 \\
0 & 0 & 0 & 1 
\end{pmatrix}. 
\end{equation}
As a result, the phase space evolution of the kicked rotor over one time period is given by the matrix 
\begin{equation}
M^{\pm}_w = M_{\text{free}} M^{\pm}_{\text{kick}, w} = 
\begin{pmatrix}
1-F^{\pm}(w)\tau & 0 & 0 & \tau \\
0 & 1-F^{\pm}(w)\tau & \tau & 0 \\
0 & -F^{\pm}(w) & 1 & 0 \\
-F^{\pm}(w) & 0 & 0 & 1
\end{pmatrix}.
\end{equation}

Notice that the matrix $M^{\pm}_w$ can be reduced to $2\times 2$ matrix
\begin{equation}
M_0^{\pm} = \begin{pmatrix} 
1 - \tau F^{\pm} & \tau \\
-F^{\pm} & 1 
\end{pmatrix}, 
\end{equation}
where the subscript $w$ is dropped from now on. 
The evolution matrices for higher multipolar orders $n$ can be derived recursively as $M_{n}^{\pm} = M_{n-1}^{\mp}M_{n-1}^{\pm}$, which determines the time evolution of duration $2^n\tau$. For example, when $B = 0$, $n = 1$ we have 
\begin{equation}
M_1^+ = \begin{pmatrix} 
-\tau^2 F^2 - \tau F + 1 & \tau^2 F + 2\tau \\
-\tau F^2 & \tau F + 1
\end{pmatrix}, \quad 
M_1^- = \begin{pmatrix}
-\tau^2 F^2 + \tau F + 1 & -\tau^2 F + 2\tau \\
-\tau F^2 & -\tau F + 1 
\end{pmatrix},
\end{equation}
where we have denoted $F := F^+ = -F^-$.

\subsection{Stability of integrable orbits}
\label{sec.stability_orbits}
We use the method proposed in Ref.~\cite{kruscha2012biased} to analyse the stability of the elliptical orbits. Let us denote 
\begin{equation}
\bar{M}_n:=\frac{1}{2}\left(M_n^{+}+M_n^{-}\right) \text{ and } D_n:=\frac{1}{2}\left(M_n^{+}-M_n^{-}\right),
\end{equation}
such that $M_n^{\pm}=\bar{M}_n+\xi D_n$ where $\xi$ is a random variable, being either $+1$ or $-1$ with the same probability. Hence, its average vanishes $\langle \xi\rangle=0$ and the variance reads $\langle \xi^2\rangle=1$.
We note that $\det(\bar{M}_n) = 1+\mathcal{O}((\tau F)^{2n})$ for non-zero $n$, implying that the averaged map $\bar{M}_n$ is not area-preserving. We therefore define a new matrix $\Bar{M}' := {\bar{M}_n}/\sqrt{\det{\bar{M}_n}}$ so that $\det (\bar{M}^{\prime}) = 1$ and this new matrix can be used to define the area of a closed orbit in the linearized system. This orbit is generally a rotated ellipse centered around the fixed point $(0,0)$. Note, $\Bar{M}'$ also depends on the multipolar order $n$ but the following method equally applies for all $n$. For now we drop it for simplicity.
With the matrix elements $M_{ij}$ of $\bar{M}'$ its area can be defined as~\cite{lichtenberg2013regular}
\begin{equation}
\label{eq.area}
A(Q_w, P_{-w})=\frac{\pi\left[M_{12} P_{-w}^2-M_{21} Q_w^2+\left(M_{11}-M_{22}\right) Q_w P_{-w}\right]}{\sqrt{1-\left(\frac{M_{11}+M_{22}}{2}\right)^2}},
\end{equation}
which is conserved if $\bar{M}'$ is repeatedly applied.
$Q_w$ and $P_{-w}$ are generally time-dependent, and in the following, we drop $w$ for simplicity and introduce $h$ to label their time-dependence. We only focus on stroboscopic time evolution and use $(Q_h,P_h)$ to represent the trajectory at time $h2^n\tau$. 
One can use the polar angle $\phi_h$ to parametrize the points $(Q_h,P_h)$ on the rotated ellipse as
\begin{equation}
    \begin{aligned}
        Q_h = R_q\cos(\theta)\cos(\phi_h)-R_p\sin(\theta)\sin(\phi_h),\\
        P_h = R_q\sin(\theta)\cos(\phi_h)+R_p\cos(\theta)\sin(\phi_h),
    \end{aligned}
\end{equation}
where $R_{q/p}$ defines length of the major or minor axis, and $\theta$ defines the rotation angle with respect to the axis. It can be determined by 
\begin{equation}
\tan(2\theta) = -\frac{M_{11}-M_{22}}{M_{12}+M_{21}}.   
\end{equation}
One can also convert the variables back as
\begin{equation}
    \cos\phi_h = \frac{Q_h \cos\theta+P_h\sin\theta}{R_q}, \ \  \sin\phi_h = \frac{-Q_h \sin\theta+P_h\cos\theta}{R_q}. 
\end{equation}
The lengths of the major and minor axes are given by 
\begin{equation}
    R_q = \frac{\sqrt{2}r_0}{\beta}\left[\frac{M_{11}-M_{22}}{\sin2\theta}+M_{12}-M_{21}\right]^{-1/2},   
    R_p = \frac{\sqrt{2}r_0}{\beta}\left[-\frac{M_{11}-M_{22}}{\sin2\theta}+M_{12}-M_{21}\right]^{-1/2},
\end{equation}
with the constant
\begin{equation}
\beta := (1-(M_{11}+M_{22})^2/4)^{-1/4}. 
\end{equation} 
It is worth noting that for $n=1$, $R_q\sim \mathcal{O}(F^{-1}), R_p\sim \mathcal{O}(F^{0})$, so for a weak kick strength, $R_q$ can be large. This stretches the ellipse in the $Q$-direction much more strongly than in the $P$-direction.

For RMD drives where $M_n^{\pm}$ is applied stochastically, the area of the closed orbit becomes time-dependent. For a single random realization and at a certain time, this area can either expand or contract. However, if we average over many different random realizations, it generally expands.
We can quantify this expansion by first defining the ellipse's radius
$
r_h=\sqrt{{A(Q_h, P_h)}/{\pi}},
$
and calculating the expansion rate ${\Delta r_h}/{r_h}$, where $\Delta r_h := r_{h+1} - r_h$. 
We use the same metric to define the ellipse's area but now the trajectory updates stochastically as
\begin{equation}
\label{eq.update_rule}
\left(\begin{array}{l}
Q_{h+1} \\
P_{h+1}
\end{array}\right)=M_n^{\pm}(w)
\left(\begin{array}{l}
Q_{h} \\
P_{h}
\end{array}\right),
\end{equation}
The expansion rate now reads 
\begin{equation}
\begin{aligned}
\label{eq.delta_r_definition}
    \frac{\Delta r_h}{r_h} := \frac{r_{h+1}}{r_h}-1 = \sqrt{\frac{M_{12} P_{h+1}^2-M_{21} Q_{h+1}^2+\left(M_{11}-M_{22}\right) Q_{h+1} P_{h+1}}{M_{12} P_h^2-M_{21} Q_h^2+\left(M_{11}-M_{22}\right) Q_h P_h}}-1.
    \end{aligned}
    \end{equation}
We now insert Eq.~\ref{eq.update_rule} into Eq.~\ref{eq.delta_r_definition} to obtain the general expression as a function of the polar angle $\phi_{h}$, the kick strength $F$ and the kick duration $\tau$. Unfortunately, it is usually very complicated and not enlightening. However, if the kick strength $F$ is small, one can perform a Taylor expansion to obtain the most relevant contributions. For our purpose, a Taylor expansion up to the order $\mathcal{O}(F^{2n})$ would be sufficient. This process can be done by employing symbolic computation tools such as \textit{Wolfram Mathematica}.

For $n=1$, to the second order of $F$, we have 
\begin{equation}
\begin{aligned}
\label{eq.delta_r_n1}
    \frac{\Delta r_h}{r_h} \approx \left(\tau F\cos 2\phi_h + \tau^2 F^2 \frac{\sqrt{2}}{2}\sin 2\phi_h\right)\xi + \left(1+\xi^2(1-\cos^22\phi_h)\right)\frac{\tau^2 F^2}{2}.
\end{aligned}
\end{equation}
Terms that are linear in $\xi$ vanish after averaging over different random realizations. Contributions that are quadratic in $\xi$ generally do not vanish, unless for special polar angles, e.g., $\cos^2 2\phi_h=1$. Also note that there is a term of order $\mathcal{O}(F)$ which does not depend on $\xi$. It arises from the fact that the average map $\bar{M}_n$ is not area-preserving, contributing a constant expansion rate as shown in Fig.~\ref{fig-linear} (the magenta line).

To further remove the angular dependence in the expansion rate, one can assume that the change in the radial direction is much slower than in the angular direction and integrate the angle $\phi_h$ over the full range $[0, 2\pi]$. We further use $\langle \xi \rangle = 0$ and $\langle \xi^2 \rangle = 1$ to obtain the averaged expansion rate that is proportional to $F^2$
\begin{equation}
\langle \frac{\Delta r_h}{r_h} \rangle = \frac{1}{2\pi} \int_{0}^{2\pi} \frac{\Delta r_h}{r_h} d\phi_h 
\approx \frac{1}{4}\tau^2 F^2 \left(2 + \langle \xi^2 \rangle \right) = \frac{3}{4}\tau^2 F^2.
\end{equation}
This provides an approximation for the mean radius evolution for small $h$ (and small $\tau F$): $\langle r_h \rangle \approx r_0 (1 + \frac{3}{4}\tau^2 F^2 h)$.

Similarly for $n=2$
\begin{equation}
\frac{\Delta r_h}{r_h} \approx \left( \left( \frac{37\sqrt{2}}{8}\tau^2 F^2 - 2\sqrt{2}\right) \tau^2 F^2 \sin{2\phi_h} - 6\tau^3 F^3 \cos{2\phi_h} \right) \xi + \left( 2\xi^2\cos{4\phi_h} + 2\xi^2 + 4\right)\tau^4 F^4, 
\end{equation}
and 
$
\langle \frac{\Delta r_h}{r_h} \rangle \approx 2\left(2 + \langle \xi^2 \rangle \right) \tau^4 F^4 = 6\tau^4 F^4.
$

Lastly for $n=3$, to the sixth order of $F$,
\begin{equation}
\begin{aligned}
\frac{\Delta r_h}{r_h} \approx & \left( -16\tau^3 F^3 \cos{2\phi_h} + 56\sqrt{2}\tau^4 F^4\sin{2\phi_h} + 206\tau^5 F^5\cos{2\phi_h} - \frac{429\sqrt{2}}{2}\tau^6 F^6\sin{2\phi_h} - 2\sqrt{2}\tau^4 F^6\sin{4\phi_h}\right) \xi \\
& + \left( -\xi^2 \cos{4\phi_h} + \xi^2 + 2\right) 64\tau^6 F^6,
\end{aligned}
\end{equation}
and the average 
$
\langle \frac{\Delta r_h}{r_h} \rangle \approx 64\left( 2 + \langle \xi^2\rangle \right) \tau^6 F^6 = 192\tau^6 F^6.
$

\subsection{Eigenvalues of the matrices $\bar{M}_n$ and $D_n$}
\label{sec.evalues}
Instead of rigorously calculating the expansion rates and their dependence on $n$, one can also estimate them by studying the scaling of eigenvalue properties of the update matrix. We first assume $B=0$ and then obtain $\bar{M}_n$ recursively as
\begin{equation}
\begin{split}
\bar{M}_0 &= \begin{pmatrix} 
1 & \tau \\
0 & 1 
\end{pmatrix}, \\
\bar{M}_1 &= \begin{pmatrix} 
1 - \tau^2 F^2 & 2\tau \\
-\tau F^2 & 1 
\end{pmatrix}, \\
\bar{M}_2 &= \begin{pmatrix} 
1 - 5\tau^2 F^2 + \tau^4 F^4 & 2\tau (2 - \tau^2 F^2) \\
\tau F^2 (-2 + \tau^2 F^2) & 1 - 3\tau^2 F^2 
\end{pmatrix}, \\
\bar{M}_3 &= \begin{pmatrix} 
1 - 18\tau^2 F^2 + 27\tau^4 F^4 - 11\tau^6 F^6 + \tau^8 F^8 & 2\tau (4 - 18\tau^2 F^2 + 10\tau^4 F^4 - \tau^6 F^6) \\
\tau F^2 (-4 + 18\tau^2 F^2 - 10\tau^4 F^4 + \tau^6 F^6) & 1 - 14\tau^2 F^2 + 9\tau^4 F^4 - \tau^6 F^6 
\end{pmatrix},
\end{split}
\end{equation}
with eigenvalues 
\begin{equation}
\begin{split}
\bar{\lambda}_{0, \pm} &= 1, \\
\bar{\lambda}_{1, \pm} &= \left(1 - \frac{1}{2}\tau^2 F^2\right) \pm \left( -\frac{1}{2}\tau F\right) \sqrt{-8 + \tau^2 F^2}, \\
\bar{\lambda}_{2, \pm} &= \left(1 - 4\tau^2 F^2 + \frac{1}{2}\tau^4 F^4\right) \pm \left(-\tau F + \frac{1}{2}\tau^3 F^3\right) \sqrt{-8 + \tau^2 F^2}, \\
\bar{\lambda}_{3, \pm} &= \left(1 - 16\tau^2 F^2 + 18\tau^4 F^4 - 6\tau^6 F^6 + \frac{1}{2}\tau^8 F^8\right) \pm \left( -2\tau F + 9\tau^3 F^3 - 5\tau^5 F^5 + \frac{1}{2}\tau^7 F^7\right) \sqrt{-8 + \tau^2 F^2}.
\end{split}
\end{equation}
Since $\tau^2 F^2 \ll 8$ the eigenvalues of $\bar{M}_n$ for each $n>0$ form a complex conjugate pair. Furthermore, as $\det (\bar{M}_n) > 1$, the deviation from one determines the rate of constant expansion of the dynamics generated by $\bar{M}_n$. Specifically, we calculate $\sqrt{\det (\bar{M}_n)}$ for different values of $n$, which correspond to the norms of the eigenvalues of $M_n$, 
\begin{equation}
\begin{split}
|\bar{\lambda}_{1, \pm}| &= 1 + \frac{1}{2}(\tau F)^2 - \frac{1}{8}(\tau F)^4 + \mathcal{O}((\tau F)^6), \\
|\bar{\lambda}_{2, \pm}| &= 1 + 4(\tau F)^4 - \frac{1}{2}(\tau F)^6 + \mathcal{O}((\tau F)^8), \\
|\bar{\lambda}_{3, \pm}| &= 1 + 128(\tau F)^6 - 160(\tau F)^8 + \mathcal{O}((\tau F)^{10}). 
\end{split}
\end{equation}
Note that for a weak kick strength, we have $(|\bar{\lambda}_{n, \pm}| - 1) \sim (\tau F)^{2n}$ for a small value of $n$, and higher-order terms are negligible. Therefore, we expect the constant expansion rate to scale as $F^{2n}$ as discussed in the previous section Sec.~\ref{sec.stability_orbits}.
However, we also observe that the prefactor for higher powers of $F$ tends to increase for $n=3$, suggesting that the perturbative expansion in orders of $F$ may not converge for a large value of $n$.

We also compute the matrix $D_n$ 
\begin{equation}
\begin{split} 
D_0 &= \begin{pmatrix} 
-\tau F & 0 \\
-F & 0 
\end{pmatrix}, \\
D_1 &= \begin{pmatrix} 
\tau F & -\tau^2 F \\
0 & -\tau F 
\end{pmatrix}, \\
D_2 &= \begin{pmatrix} 
\tau^3 F^3 & \tau^2 F (4 - \tau^2 F^2) \\
2\tau^2 F^3 & -\tau^3 F^3 
\end{pmatrix}, \\
D_3 &= \begin{pmatrix} 
-16\tau^3 F^3 + 10\tau^5 F^5 - \tau^7 F^7 & \tau^4 F^3 (16 - 10\tau^2 F^2 + \tau^4 F^4) \\
0 & 16\tau^3 F^3 - 10\tau^5 F^5 + \tau^7 F^7
\end{pmatrix};
\end{split}
\end{equation}
with eigenvalues
\begin{equation}
\begin{split} 
\mu_{0, \pm} &= 0, -\tau F, \\
\mu_{1, \pm} &= \mp \tau F, \\
\mu_{2, \pm} &= \pm \tau^2 F^2 \sqrt{8 - \tau^2 F^2}, \\
\mu_{3, \pm} &= \pm (16\tau^3 F^3 - 10\tau^5 F^5 + \tau^7 F^7),
\end{split}
\end{equation}
which scales as $|\mu_{n, \pm}| \sim (\tau F)^n$ for $n>0$. As $D_n$ appears stochastically in time, we expect its leading-order contribution to vanish. Its second-order effects lead to a diffusive spiral-out process with an expansion rate that scales as $F^{2n}$.

We perform a similar calculation for non-zero $B$ and by defining $F^{\pm} \sim (\pm K + B)$, we obtain
\begin{equation}
\begin{split}
|\bar{\lambda}_{1, \pm}| &= \sqrt{1 + \frac{\tau^4}{4}(F^+ - F^-)^2} = 1 + \frac{\tau^2}{8}(F^+ - F^-)^2 + \mathcal{O}((F^+ - F^-)^3), \\
|\bar{\lambda}_{2, \pm}| &= 1 + \frac{\tau^3}{8}(F^+ - F^-)^2 \left(8 - 3\tau (F^+ + F^-) + \tau^2 F^+ F^-\right) \left(F^+ + F^- - \tau F^+ F^-\right) + \mathcal{O}((F^+ - F^-)^3), \\
|\bar{\lambda}_{3, \pm}| &= 1 + \frac{\tau^4}{8}(F^+ - F^-)^2 (F^+ + F^- - \tau F^+ F^-)^2 \left( 8 - 3\tau (F^+ + F^-) + \tau^2 F^+ F^-\right)^2 \left( 2 - 2\tau (F^+ + F^-) + \tau^2 F^+ F^-\right)^2 \\
&\quad + \mathcal{O}((F^+ - F^-)^3). 
\end{split}
\end{equation}
Importantly, one always finds $(|\bar{\lambda}_{n, \pm} - 1|) \sim K^2$, where the scaling exponent does not depend on the multipolar order. This scaling relation corresponds to the observed heating rate scaling of $K^2$ in Fig.~\ref{fig-B0pt01} in the main text.

\section{Temperature at the prethermal stage} 
\label{sec.temperature}

\begin{figure}
\centering
\includegraphics[width=0.9\linewidth]{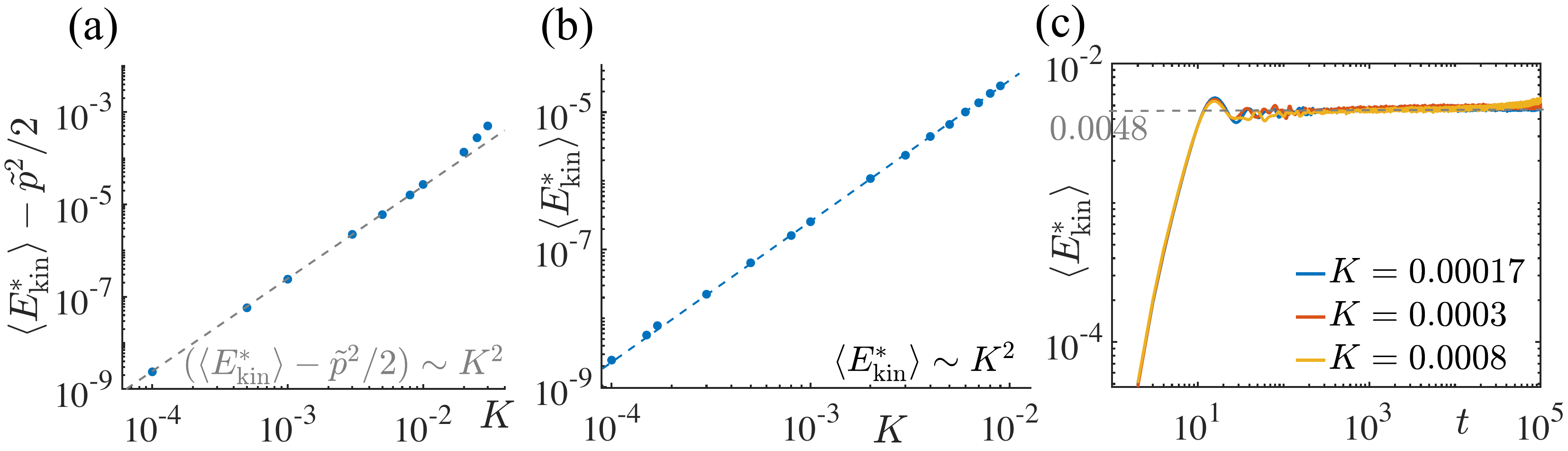}
\caption{The averaged prethermal kinetic energy density $\langle E_{\text{kin}}^*\rangle$ for $n=1$ RMD for (a) $B = 0$ with $\tilde{p} = 0.1$, (b) $B = 0$ with $\tilde{p} = 0$ and (c) $B = 0.01$ with $\tilde{p} = 0$. The prethermal temperature scales as $T\sim K^2$ for (a) and (b) when the kick is weak. In the last case it is fixed by $B$, so all the plateau parts of energy curves are at the same value $\langle E_{\text{kin}}^*\rangle = 0.0048$. The rest numerical settings are the same as in Fig.\ref{fig-B0}. All plots use a log-log scale.}
\label{fig.temperature}
\end{figure}
Here we analyse the dependence of the prethermal temperature on the kick strength and we   demonstrate that it follows $T\sim K^2$.  For the initial condition $p_j=\tilde{p}$, the initial kinetic energy density is given by $\tilde{p}^2/2$. In the prethermal regime with a weak kick, we assume that the distribution for $p_j$ and $q_j$ decouples~\cite{rajak2019characterizations}. The angular momentum distribution approaches the
Gibbs distribution 
$Z^{-1}\prod_{j=1}^N \exp \left[-{(p_j-\tilde{p})^2}/{2 T}\right]$ with a normalization factor $Z^{-1}$. The corresponding kinetic energy density is $E^{*}_{\mathrm{kin}}=(T+\tilde{p}^2)/2$. We numerically study the dependence of the temperature $T$ on the kick strength $K$ in Fig.~\ref{fig.temperature} (a) and (b) for two different initial conditions: $\tilde{p}=0.1$ and $\tilde{p}=0$, respectively. The numerical results fit well with a straight line in a log-log scale, with a slope around 2. This suggests that the temperature follows $T\sim K^2$. 

In contrast, for the modified kicked protocol used in Fig.~\ref{fig-B0pt01} with a non-vanishing $B$, the averaged prethermal kinetic energy linearly depends on $B$ but does not notably change with $K$. We verify this in Fig.~\ref{fig.temperature}, panel (c).

\section{Distribution of angular momenta}
\label{sec.momentum_distribution}
In the main text, we linearize the many-body Hamiltonian and explain the characteristic scaling of the prethermal lifetime. We note that such scaling can be very stable and persist even away from the linearization regime $(q_j - q_{j+1}) \text{ mod } 2\pi \ll 1$, as shown in Fig.~\ref{fig-B0} where a wide initial distribution of $q_j$ is used. Interestingly, we notice that this lifetime scaling is not sensitive to the angular dependence, but strongly relies on the angular momentum distribution during the prethermal regime. As discussed in Sec.~\ref{sec.temperature}, this distribution is governed by the prethermal temperature. We find that, as long as the prethermal regime exhibits a low temperature, or equivalently, a narrow distribution of angular momenta, the prethermal regime can be sufficiently long-lived, and the dependence on $n$ should manifest. 

\begin{figure}[h]
\centering
\includegraphics[width=0.6\linewidth]{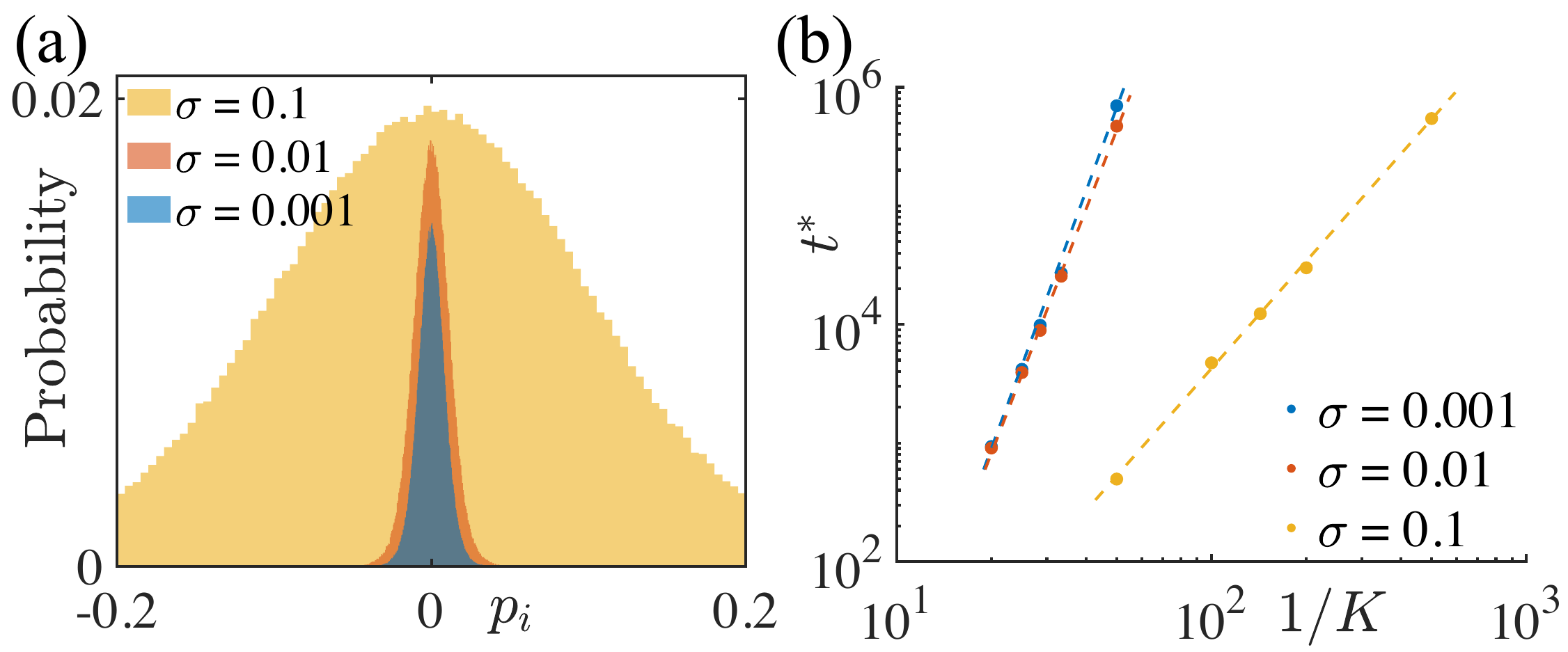}
\caption{(a)
Momentum distribution at the end of the prethermal stage for a Gaussian initial momentum distribution with a zero mean and a standard deviation $\sigma$ indicated in different colours, and (b) corresponding prethermal lifetime $t^*$ as a function of $1/K$ for $n=20$ in log-log. The rest numerical settings are the same as in Fig.\ref{fig-B0}. 
}
\label{fig.momentum_distribution}
\end{figure}

The prethermal temperature can be adjusted by the initial condition. For instance, we consider initial an angular momentum distribution following a Gaussian distribution with a zero mean and a standard deviation $\sigma$. A larger standard deviation generally increases the prethermal temperature, resulting in a broader angular momentum distribution during the prethermal regime. This is confirmed in Fig.~\ref{fig.momentum_distribution}, panel (a), where three different values of the initial standard deviation $\sigma$ are used. Note that we use $n=1$ to generate the dynamics but this figure qualitatively represents other multipolar order as well. The kick strength $K$ is chosen such that the prethermal lifetimes are approximately the same for all $\sigma$, with values of $K$ set as $0.012$, $0.008$ and $0.005$ for $\sigma=0.001$, $0.01$ and $0.1$, respectively. The probability distributions are extracted at $t=1500$ just before the system notably heats up. 

We now illustrate the dependence of the prethermal lifetime scaling on different initial conditions, focusing on the TM drive. For a fixed kick strength $K$, it typically determines the longest possible prethermal lifetime for the entire family of $n$-RMD protocols. Therefore, if we fit $t^*$ versus $1/K$ on a log-log scale, the scaling exponent also sets the upper bound for other $n$-RMD protocols. As long as the TM drive exhibits a sufficiently large scaling exponent $\alpha$, $n$-RMD with any finite $n$ should exhibit the $n$-dependence in the lifetime scaling.

In Fig.~\ref{fig.momentum_distribution}, panel (b), we present the prethermal lifetime scaling for different initial conditions. For narrow distributions, such as $\sigma=0.001$ and $0.01$, the scaling exponents are still very large, approximately $\alpha\approx 7.9$ and $7.2$, respectively. We expect that these fitted scaling exponents may increases further if we perform the fit using larger $1/K$ and longer time windows, similar to Fig.~\ref{fig-B0}(c). However, for a larger standard deviation, such as $\sigma=0.1$, the scaling exponent notably decreases to $\alpha\approx3.1$, and we expect that $n$-RMD systems with finite $n$ would heat up faster. 

Therefore, we expect that as long as the prethermal regime has a low temperature, a long-lived prethermal regime and the $n$-dependence in the lifetime scaling should emerge. Further systematic investigations of the temperature dependence of the prethermal lifetime scaling will be explored in future work.

\section{Scaling of the Thue-Morse prethermal lifetime}
\label{sec.tm}
\begin{figure}[h]
\centering
\includegraphics[width=0.6\linewidth]{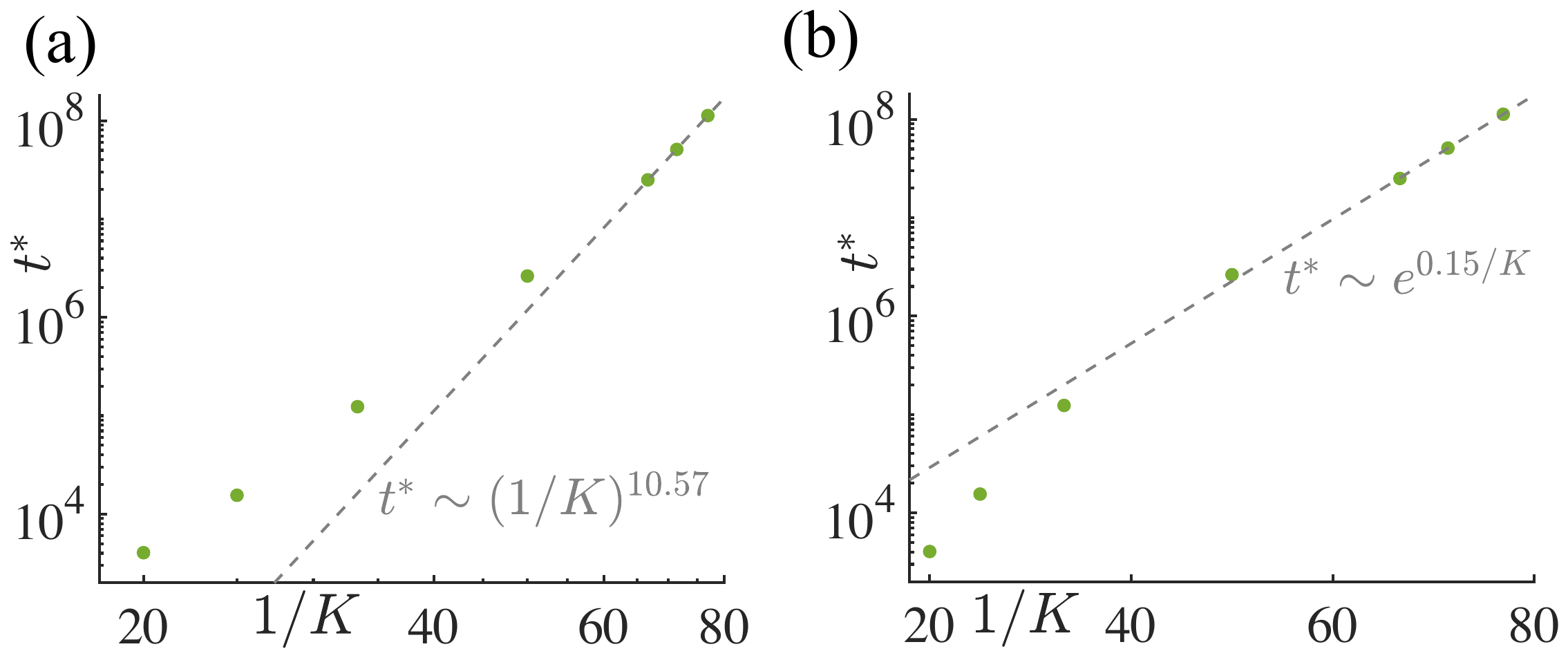}
\caption{ (a) and (b) depict the power-law and exponential fitting of the prethermal lifetime $t^*(1/K)$ for the TM drive.}
\label{fig.TM_scaling}
\end{figure}

In the main text we show that for the TM drive, the lifetime scaling becomes $t^*\sim \exp(C[\ln(K^{-1}/g)]^2)$ with constants $C$ and $g$. Here, we compare this result with other fitting methods. For instance, in Fig.~\ref{fig.TM_scaling}(a) we use a log-log scale and clearly the numerical data tends to curve up. In contrast, panel (b) depicts the same data but in log scale and the numerical result bends down.
Therefore, this scaling grows faster than any power-law but slower than exponentially.

\subsection{Finite size effects}
\label{sec.finite_size}
\begin{figure}[h]
\centering
\includegraphics[width=0.3\linewidth]{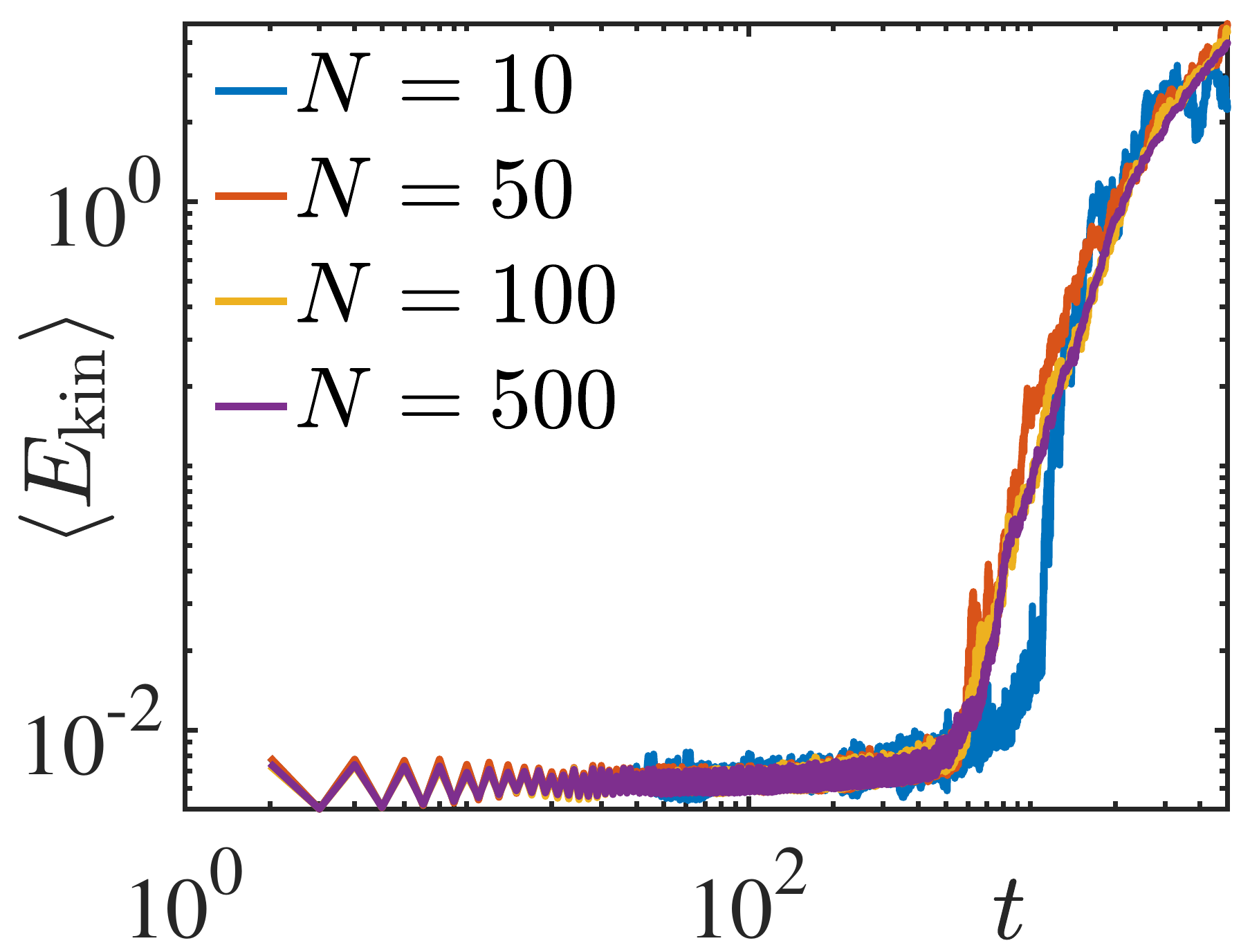}
\caption{Time evolution of the averaged kinetic energy for the TM drive. The simulation results converge for large systems and $N=500$ is already sufficient to produce thermodynamically large systems. 
Here we use the kick strength $K=0.07$ and the results are averaged over 200 random realizations. Initial states are the same as in Fig.~\ref{fig-B0}.}
\label{fig.finite_size}
\end{figure}

In Fig.~\ref{fig.finite_size} we compare the dynamics using the TM drive for different system sizes. The simulations converge as the system size increases. In the main text, we use $N=500$ to generate the data, which is already sufficient to mimic the heating behaviour in thermodynamically large systems.

\end{document}